\input harvmac.tex 
\input amssym.def
\input amssym
\input epsf.tex
\magnification\magstep1
\baselineskip 14pt

\parskip 6pt
\newdimen\itemindent \itemindent=32pt
\def\textindent#1{\parindent=\itemindent\let\par=\resetpar%
\indent\llap{#1\enspace}\ignorespaces}

\let\oldpar=\par
\def\resetpar{\oldpar\parindent=20pt\let\par=\oldpar}

\font\ninerm=cmr9 \font\ninesy=cmsy9
\font\eightrm=cmr8 \font\sixrm=cmr6
\font\eighti=cmmi8 \font\sixi=cmmi6
\font\eightsy=cmsy8 \font\sixsy=cmsy6
\font\eightbf=cmbx8 \font\sixbf=cmbx6
\font\eightit=cmti8
\def\eightpoint{\def\rm{\fam0\eightrm}
 \textfont0=\eightrm \scriptfont0=\sixrm \scriptscriptfont0=\fiverm
 \textfont1=\eighti  \scriptfont1=\sixi  \scriptscriptfont1=\fivei
 \textfont2=\eightsy \scriptfont2=\sixsy \scriptscriptfont2=\fivesy
 \textfont3=\tenex   \scriptfont3=\tenex \scriptscriptfont3=\tenex
 \textfont\itfam=\eightit  \def\it{\fam\itfam\eightit}%
 \textfont\bffam=\eightbf  \scriptfont\bffam=\sixbf
 \scriptscriptfont\bffam=\fivebf  \def\bf{\fam\bffam\eightbf}%
 \normalbaselineskip=9pt
 \setbox\strutbox=\hbox{\vrule height7pt depth2pt width0pt}%
 \let\big=\eightbig  \normalbaselines\rm}
\catcode`@=11 %
\def\eightbig#1{{\hbox{$\textfont0=\ninerm\textfont2=\ninesy
 \left#1\vbox to6.5pt{}\right.\n@@space$}}}
\def\vfootnote#1{\insert\footins\bgroup\eightpoint
 \interlinepenalty=\interfootnotelinepenalty
 \splittopskip=\ht\strutbox %
 \splitmaxdepth=\dp\strutbox %
 \leftskip=0pt \rightskip=0pt \spaceskip=0pt \xspaceskip=0pt
 \textindent{#1}\footstrut\futurelet\next\fo@t}
\catcode`@=12 %
\def \de{\delta}
\def \De{\Delta}
\def \si{\sigma}

\def \pr{\partial}
\def \tr{{\rm tr }}

\def \l{\big \langle}
\def \r{\big \rangle}

\def \half{{\textstyle {1 \over 2}}}

\def \ts{\textstyle}

\def \b{{\rm b}}
\def \d{{\rm d}}

\def \x{{\rm x}}
\def \y{{\rm y}}

\def \C{{\cal C}}

\def \E{{\cal E}}
\def \F{{\cal F}}

\def \M{{\cal M}}
\def \N{{\cal N}}

\def \P{{\cal P}}

\def \S{{\cal S}}

\def \Z{{\cal Z}}

\def \bj{{\bar \jmath}}

\def \b{{\rm b}}
\def \e{{\rm e}}
\def \x{{\rm x}}
\def \y{{\rm y}}
\def \t{{\rm t}}
\def \z{{\rm z}}

\def \of{{\overline f}}

\def \al{{\alpha}}
\def \tr{{\rm tr}}
\def \u{{\rm u}}
\def \be{\beta}
\def \dps{\displaystyle}

\font \bigbf=cmbx10 scaled \magstep1

\lref\ska{ B.-S. Skagerstam, {\it On the Large $N_c$ Limit of the $SU(N_c)$
Colour Quark-Gluon Partition Function}, Z. Phys. C 24 (1984) 97.}

\lref\sund{B. Sundborg,
 {\it The Hagedorn Transition, Deconfinement and $N = 4$ SYM Theory},
 Nucl.  Phys.  B 573 (2000) 349, hep-th/9908001.}

\lref\raam{ O. Aharony, J. Marsano, S. Minwalla, K. Papadodimas and M. Van
Raamsdonk,
 {\it The Hagedorn/Deconfinement Phase Transition in Weakly Coupled Large 
$N$
 Gauge Theories},
 Adv.  Theor.  Math.  Phys.  8 (2004) 603, hep-th/0310285.}

\lref\mac{I.G. Macdonald, {\it Symmetric Functions and Hall Polynomials},
Second Edition, Clarendon Press, Oxford, 1995.}

\lref\char{M. Bianchi, F.A. Dolan, P.J. Heslop and H. Osborn,
{\it $\N=4$ Superconformal Characters and Partition Functions},
Nucl. Phys. B 767 [FS] (2007) 163, hep-th/0609179.}

\lref\mald{
J.~Kinney, J.M.~Maldacena, S.~Minwalla and S.~Raju,
{\it An Index for $4$ Dimensional Super Conformal Theories},
hep-th/0510251.}

\lref\poly{
A.M. Polyakov,
 {\it Gauge Fields and Space-Time},
 Int. J. Mod. Phys.  A 17S1 (2002) 119,
 hep-th/0110196.}

\lref\sag{B.E. Sagan, {\it The Symmetric Group, Representations,
Combinatorial Algorithms, and Symmetric Functions}, Wadsworth and
Brooks/Cole Mathematics Series, California, 1991.}

\lref\gess{I.M. Gessel, {\it Symmetric Functions and P-Recursiveness}, T.
Combin. Theory Ser. A 53 (1990) 257.}

\lref\carl{J. Carlsson and B.H.J. McKellar,
 {\it $SU(N)$ Glueball Masses in $2+1$ Dimensions},
 Phys. Rev.  D 68 (2003) 074502, hep-lat/0303016.}

\lref\cumm{L. Begin, C. Cummins and P. Mathieu,
 {\it Generating Functions for Tensor Products},
 hep-th/9811113.}

\lref\andrews{G.E. Andrews, R. Askey and R. Roy, {\it Special Functions},
Cambridge University Press, Cambridge, 1999.}

\lref\stan{R.P. Stanley, {\it Enumerative Combinatorics}, Vol. 2, Cambridge
University Press, Cambridge, 1999.}

\lref\odl{A.M. Odlyzko, {\it Asymptotic Enumeration Methods}, Handbook of
Combinatorics,
   Vol. 2, eds. R.L. Graham, M. Gr{\"o}tschel and L. Lov\'asz,
   North-Holland,
   Amsterdam, 1995.}

\lref\bofeng{S. Benvenuti, B. Feng, A. Hanany, Y.H. He,
 {\it Counting $BPS$ Operators in Gauge Theories: Quivers, Syzygies and
 Plethystics}, hep-th/0608050.}

\lref\bofengt{B. Feng, A. Hanany, Y.H. He,
 {\it Counting Gauge Invariants: The Plethystic Program}, hep-th/0701063.}

\lref\nak{Y. Nakayama, {\it Finite $N$ Index and Angular Momentum Bound from
Gravity}, hep-th/0701208.}

\lref\willen{J.F. Willenbring,
 {\it Stable Hilbert Series of $\S({\frak g})^K$ for Classical Groups}, 
math/0510649}

\lref\maldo{J.M. Maldacena,
 {\it The Large $N$ Limit of Superconformal Field Theories and 
Supergravity},
 Adv. Theor. Math. Phys.  2 (1998) 231, hep-th/9711200.}

\lref\dhok{E. D'Hoker, P. Heslop, P. Howe and A.V. Ryzhov,
 {\it Systematics of Quarter BPS Operators in $N = 4$ SYM},
 JHEP 0304 (2003) 038, hep-th/0301104.}

\lref\gea{G.E. Andrews, {\it The Theory of Partitions}, Encyclopedia of
Mathematics and its Applications, Vol. 2, ed. G.C. Rota, Addison-Wesley,
Reading, Massachusetts, 1976.}

\lref\mart{D. Martelli, J. Sparks and S.T. Yau, {\it Sasaki-Einstein
Manifolds and Volume Minimisation}, hep-th/0603021.}

\lref\ray{G. Mandal and N.V. Suryanarayana,
 {\it Counting $1/8$ BPS Dual Giants}, 
hep-th/0606088.}

\lref\sinha{A. Sinha and N.V. Suryanarayana,
 {\it Two-charge Small Black Hole Entropy: String-loops and 
Multi-strings},
 JHEP 0610 (2006) 034, hep-th/0606218.}

\lref\marsp{D. Martelli and J. Sparks, {\it Dual Giant Gravitons in
Sasaki-Einstein Backgrounds}, Nucl. Phys. B 759 (2006) 292, hep-th/0608060.}

\lref\basu{A. Basu and G. Mandal, {\it Dual Giant Gravitons in ${\rm
AdS}_m\times Y^n$ (Sasaki-Einstein)}, hep-th/0608093}

\lref\butti{A. Butti, D. Forcella and A. Zaffaroni, {\it Counting BPS
Baryonic Operators in CFTs with Sasaki-Einstein Duals}, hep-th/0611229.}

\lref\hanany{A. Hanany and C. Romelsberger, {\it Counting BPS Operators in
the Chiral Ring of $N = 2$ Supersymmetric Gauge
 Theories or $N = 2$ Braine Surgery}, hep-th/0611346.}

\lref\grant{L. Grant and K. Narayan, {\it Mesonic Chiral Rings in Calibi-Yau
Cones from Field Theory}, hep-th/0701189.}

\lref\force{D. Forcella, A. Hanany and A. Zaffaroni, {\it Baryonic
Generating Functions},
hep-th/0701236.}

\lref\lee{S. Lee, S. Lee and J. Park, {\it Toric ${\rm AdS}_4/{\rm CFT}_3$
Duals and M-theory Crystals}, hep-th/0702120.}

\lref\sury{N.V. Suryanarayana,
 {\it Half-BPS Giants, Free Fermions and Microstates of Superstars},
 JHEP 0601 (2006) 082, hep-th/0411145.}

\lref\gut{J.B. Gutowski and H.S. Reall,
 {\it General Supersymmetric AdS$(5)$ Black Holes},
 JHEP 0404 (2004) 048, hep-th/0401129.}

\lref\guto{J.B. Gutowski and H.S. Reall, {\it Supersymmetric AdS$(5)$ Black
Holes}, JHEP 0402 (2004) 006, hep-th/0401042.}

\lref\chong{Z.W. Chong, M. Cvetic, H. Lu and C.N. Pope, {\it General
Non-extremal Rotating Black Holes in Minimal Five-Dimensional Gauged
Supergravity}, Phys. Rev. Lett. 95 (2005) 161301 hep-th/0506029.}

\lref\dob{V.K. Dobrev,
 {\it Characters of the Positive Energy UIRs of $D = 4$ Conformal 
Supersymmetry},
 hep-th/0406154.}

\lref\dobsez{V.K. Dobrev and E. Sezgin, {\it Spectrum and Character Formulae
of $SO(3,2)$ Unitary Representations}, Lecture Notes in Physics, Vol. 379,
eds. J.D Hennig, W. L\"ucke and J. Tolar, Springer-Verlag, Berlin, 1990.}

\lref\dobt{V.K. Dobrev, {\it Positive Energy Representations of Non-compact
Quantum Algebras}, Proceedings of the Workshop on Generalized Symmetries in
Physics, Clausthal, July 1993, eds. H.D. Doebner et al., World Sci.
Singapore, 1994.}

\lref\barab{A. Barabanschikov, L. Grant, L.L. Huang and S. Raju, {\it The
Spectrum of Yang Mills on a Sphere}, hep-th/0501063.}

\lref\mep{F.A. Dolan, {\it Character Formulae and Partition Functions in
Higher Dimensional Conformal Field Theory},
 J. Math. Phys. 47 (2006) 062303, hep-th/0508031.}

\lref\gib{G.W. Gibbons, M.J. Perry and C.N. Pope,
 {\it Partition Functions, the Bekenstein Bound and Temperature Inversion 
in
 anti-de Sitter Space and its Conformal Boundary},
 Phys. Rev.  D 74 (2006) 084009, hep-th/0606186.}

\lref\cardy{J.L. Cardy, {\it Operator Content and Modular Properties of
Higher Dimensional Conformal Field Theories}, Nucl. Phys. B366 (1991) 403.}

\lref\kut{D. Kutasov and F. Larsen,
 {\it Partition Sums and Entropy Bounds in Weakly Coupled CFT},
JHEP 0101 (2001) 001, hep-th/0009244.}

\lref\lit{D.E. Littlewood, {\it The Theory of Group Characters and Matrix
Representations of Groups}, Clarendon Press, Oxford, 1940.}

\lref\wright{E.M. Wright, {\it Asymptotic Partition Formulae, I:
Plane Partitions}, Quart. J. Math. Oxford Ser.  2 (1931) pp. 177-189.}

\lref\Lin{
  H. Lin and J.M. Maldacena,
  {\it Fivebranes from Gauge Theory},
  Phys. Rev.  D 74 (2006) 084014, hep-th/0509235.}

\lref\ono{E. Onofri, G. Veneziano and J. Wosiek, {\it Supersymmetry
and Combinatorics},  math-ph/0603082.}

\lref\pietri{R. De Pietri, S. Mori and E. Onofri, {\it The Planar Spectrum in
$U(N)$-Invariant Quantum Mechanics by Fock Space Methods: I. The Bosonic
Case}, JHEP 0701 (2007) 018, hep-th/0610045.}

\lref\bonin{M. Bonini, G.M. Cicuta and E. Onofri, {\it Fock Space Methods and Large
$N$}, J. Phys. A  40 (2007) F229, hep-th/0701076.}

\lref\luc{H.K. Kunduri, J. Lucietti and
H.S. Reall, {\it Supersymmetric Multi-Charge AdS$(5)$ Black Holes}, JHEP 0604 (2006) 036, hep-th/0601156.}

\lref\luct{
  H.K. Kunduri, J. Lucietti and H.S. Reall,
  {\it Do Supersymmetric Anti-de Sitter Black Rings Exist?},
  JHEP 0702 (2007) 026, hep-th/0611351.}

{\nopagenumbers
\rightline{DIAS-STP-07-05}
\rightline{arXiv:0704.1038 [hep-th]}
\vskip 1.5truecm \centerline {\bigbf  Counting BPS Operators in $\N=4$ SYM}
\vskip  6pt
\vskip 2.0 true cm
\centerline {F.A. Dolan}

\vskip 12pt
\centerline {Institi\'uid Ard-l\'einn Bhaile \'Atha Cliath,}
\centerline {(Dublin Institute for Advanced Studies,)}
\centerline {10 Burlington Rd., Dublin 4, Ireland}
\vskip 1.5 true cm

{\eightpoint
\parindent 1.5cm{
\noindent
{\narrower\smallskip\parindent 0pt

The free field partition function for a generic $U(N)$ gauge theory, where
the fundamental fields transform in the adjoint representation, is analysed
in terms of symmetric polynomial techniques. It is shown by these means how
this is related to the cycle polynomial for the symmetric group and how the
large $N$ result may be easily recovered. Higher order corrections for
finite $N$ are also discussed in terms of symmetric group characters. For
finite $N$, the partition function involving a single bosonic fundamental
field is recovered and explicit counting of
 multi-trace quarter BPS operators in free $\N=4$ super Yang Mills
discussed, including a general result for large $N$.
The partition function for quarter BPS operators in the chiral ring of
$\N=4$ super Yang Mills is analysed in terms of plane partitions.
Asymptotic counting of BPS primary operators with
differing $R$-symmetry charges is discussed in both free $\N=4$ super
Yang Mills and in the chiral ring.
Also, general and explicit expressions are derived for $SU(2)$ gauge theory
partition functions, when the fundamental fields transform in the
adjoint,
 for free field theory.

Keywords:
Characters, Partition Functions, Gauge Theory, $\N=4$ Super Yang Mills

\narrower}}

\vfill
\line{\hskip0.2cm E-mail:
{{\tt fdolan@stp.dias.ie}}\hfil}
}

\eject}
\pageno=1

\noindent
\newsec{Introduction}

For the past while there has been intense interest in finite $N$ partition
functions for Yang Mills theories, especially in super-symmetric ones,
particularly with regard to their construction for BPS states and the
counting thereof \refs{\mald, \mart, \bofeng, \marsp, \basu, \butti,
\hanany, \grant, \bofengt, \force, \lee,\ray, \sinha, \sury, \barab}. Much
attention has been devoted to this issue for $\N=4$ super Yang
Mills theory, it being by now the archetypal example of a conformal field
theory for which we have a dual description in terms of string theory, by
means of the AdS/CFT conjecture \maldo. An example is in the assiduous
efforts that have been made to explain the entropy of certain BPS black
holes in AdS$_5\times S^5$ \refs{\guto,\gut,\chong,\luc} in terms of microscopic
counting of dual operators in $\N=4$ super Yang Mills, with gauge group
$SU(N)$, by means of partition functions \mald. While this problem remains
unsolved to date, essentially due to the difficulty of defining what is
meant by these dual operators for finite $N$, there are other equally
interesting sectors of (super) Yang Mills theories where more progress with
counting has been made, examples being in free $\N=4$ super Yang Mills
and in chiral ring sectors involving BPS operators.

For $\N=4$ super Yang Mills, the half BPS sector
consists of multi-trace operators involving a single bosonic operator, $Z$.
Similarly, the quarter BPS sector consists of multi-trace operators
involving two bosonic operators $Z,Y$ while the eighth consists of ones
involving three bosonic operators $Z,Y,X$ and two fermionic operators
$\lambda,{\bar \lambda}$.  (All these operators are here assumed to belong to the Lie algebra
of $U(N)$.)

For the chiral ring, the commuting/anti-commuting of these operators is at the heart of
why we can write very concise and elegant generating
functions for the finite $N$ multi-trace partition functions \mald. Analysis
of these partition functions, in terms of the counting of operators, has
become a sophisticated industry where such approaches as the so-called
`Plethystic Program' have provided substantial results
\refs{\bofeng,\bofengt}.

This paper is devoted largely to the issue of partition functions for free
field theory and particularly to the counting of gauge invariant multi-trace
operators for the case of two bosonic fundamental fields. This is of
relevance to the quarter BPS sector of $\N=4$ super Yang Mills
in the free field limit when the operators $Z,Y$ do not commute, 
in contrast to the chiral ring when they do.

The partition function for a free, massless quark-gluon gas was computed
long ago \ska.  This
involved taking particle statistics into account using
coherent state techniques and then imposing the gauge singlet
condition by integrating over the relevant gauge group.
With some modifications to the expression thus derived
we may write the multi-trace partition function for some generic
bosonic/fermionic fundamental fields in terms of an integral
over the gauge group, involving the single particle partition function
\refs{\sund,\raam}.  This is the starting point here.

For $U(N)$ gauge theories we may easily write down the integral, though,
even for this case, its evaluation is far from simple. One approach (which
we adapt here for the $SU(2)$ case) is to rewrite the expression in terms of
an $N$-fold contour integral, whereby it may in principle be evaluated by
summing the contributions from poles inside origin centred unit discs, in each
of the $N$ complex planes - similar techniques have been used in
\force. 
Due to the number of poles this becomes
unfeasible for higher values of $N$. Another approach is to use the fact
that the complex integral that interests us provides for an inner
product for symmetric polynomials - see Macdonald \mac, pp. 363 - 372, for a
related discussion. Taking this point more seriously reveals an alternative
route to evaluating the free field theory partition function which exposes
not only the large $N$ case in an almost trivial way, but also how and
where this differs from the finite $N$ case.

This treatment also
reveals an alternative interpretation of the free field partition function
at finite $N$ - it is related to a gauge group
average of the cycle polynomial for the symmetric permutation group (after a
certain identification of `letters' with
gauge group valued variables).  This point is not
dwelt on further here though makes the connection between
the partition function and Polya enumeration
explicit.  For single trace operators at large $N$
this connection has already been made \refs{\sund,\poly}
for $\N=4$ super Yang Mills, whereby the partition
function for single trace operators is related
to the cycle polynomial for the cyclic permutation group.{\foot{
For more recent applications of symmetric polynomials and Polya enumeration
to super-symmetric quantum mechanical models, analysed in terms of
Fock space methods, see \refs{\ono, \pietri, \bonin}.  }}

Another issue is how to use the expression for
the free field partition function to give explicit counting of
gauge invariant multi-trace operators
in a Yang-Mills theory, with gauge group $U(N)$.  The case for one bosonic
fundamental field has been widely
discussed and, for finite $N$, the operators are counted
by partition numbers of non-negative integers into
at most $N$ parts (which is here denoted by $p_N(n)$
- no closed formula for these numbers for arbitrary $N, n$ exist, though
they have a `nice' generating function).  Here, this result is re-derived,
from a symmetric function perspective, by
employing the well known Cauchy-Littlewood
formula.

To proceed further with counting, for the quarter BPS sector of $\N=4$ super
Yang Mills, for instance, character methods prove to be both natural and
indispensable. Characters in relation to conformal field theories prove to
be very convenient for encoding the allowable representations \refs{\barab,
\dobsez, \dobt, \mep, \dob, \char} and for studying related partition
functions, \refs{\gib, \cardy, \kut, \char}. For $\N=4$ super Yang Mills, it
was shown in \char\ that, if we are to distinguish among 
primary operators with
differing conformal dimensions, spins and $R$-symmetry charges, such
counting is most easily achieved using reductions of the full $\N=4$
superconformal characters, in certain limits that isolate corresponding
sectors of short/semi-short operators. (One such limit corresponds to the
index constructed in \mald.) This point may be easily illustrated for
quarter BPS primary operators in $\N=4$ super Yang Mills, the case of two bosonic
fundamental fields here. (See \dhok\ for an explicit construction and
counting of quarter
BPS operators.) For $\N=4$ super Yang Mills, the counting of quarter BPS
primary operators is complicated by the fact that, if we are to keep track of
differing $R$-symmetry representations, any partition function restricted to
this sector must be expanded in terms of $U(2)$ characters (or two-variable
Schur polynomials). Denoting some partition function restricted to this
sector by $\Z(t,u)$, where $t,u$ are letters
corresponding to the fields $Z,Y$, then by expanding \eqn\exptwobosc{
\Z(t,u)=\sum_{n=0}^\infty\sum_{m=0}^n\N_{(n,m)} \,s_{(n,m)}(t,u) \, , \qquad
s_{(n,m)}(t,u)={t^{n+1}u^{m}-t^m u^{n+1}\over t-u} \, , } in terms of
two-variable Schur polynomials $s_{(n,m)}(t,u)$, we obtain the numbers
$\N_{(n,m)}$ of gauge invariant quarter BPS operators belonging to the
$[m,n{-}m,m]$ $SU(4)_R$ $R$-symmetry representation and having conformal
dimension $n+m$ (so that they are superconformal primary highest weight
states in the corresponding quarter BPS supermultiplets). (The case $m=0$
counts
gauge invariant half
BPS primary operators.)

Here, the free field partition function is thus expanded in terms of Schur
polynomials, depending on the same variables as the one particle partition
function, the two boson case being a specialisation. This is quite naturally
achieved using the Cauchy-Littlewood formula (and, if we include fermions,
another formula due to Littlewood). Generally, we may obtain a result that
relates the counting numbers to a sum over Kronecker coefficients. These
arise naturally in the theory of the symmetric permutation group, though
remain somewhat mysterious from a combinatorial perspective.

Specialising to the two boson case, a recursive procedure is employed here
for the counting of multi-trace quarter
BPS operators in free field theory at finite $N$.
This issue was given considerable discussion for the large $N$ case in
\char\ - here the results of \char\
are generalised in terms of a generating
function that may be employed to
count quarter BPS operators for any $R$-symmetry
charges, at large $N$.  Asymptotic counting is also addressed
in the latter case for the numbers $\N_{(n,m)}$ in \exptwobosc\
for large $n$ and fixed $m$.

To complete the discussion, counting of quarter BPS operators in the chiral
ring of $\N=4$ super Yang Mills is investigated in terms of expanding over
$U(2)$ characters as in \exptwobosc. An explicit formula is given for the
corresponding finite $N$ partition function, with a short combinatorial
interpretation given in terms of plane partitions, 
and specialised to large $N$. For the latter case, the
exponential behaviour of the numbers $\N_{(n,m)}$ in \exptwobosc\ is found for
$n$ and $m$ both comparably large. This behaviour is consistent with a
special case addressed in \refs{\bofeng,\bofengt}.  By way of completion,
a similar discussion for an arbitrary
number of bosonic fundamental fields in the chiral ring is included.

Two appendices are included; the first establishes some notation used for
partitions and gives some standard results for the symmetric
group  and symmetric polynomials,
the second gives some tables of numbers of quarter BPS operators
in free $\N=4$ super Yang Mills, with gauge group $U(N)$, for which
explicit formulae are given in the main text.  Footnotes contain
further details and points of clarification.

\noindent
\newsec{Free Field Partition Functions}

We start from the single particle partition function which is here denoted by
$f(\t)$ for some variables{\foot{In what follows roman letters are used to
denote a collection of variables and, for $\x=(x_1,\dots,x_i)$,
$\y=(y_1,\dots,y_j)$ for example, the shorthand $\x^\al$ is used to mean
$(x_1{}^\al,\dots,x_i{}^{\al})$ and $\z=\x \y$ to mean
$\z=(z_{11},\dots,z_{ij})$ where $z_{rs}=x_r y_s$. The latter convenient
notation has been used by Macdonald \mac.}} $\t=(t_1,t_2\dots)$.
The general form of $f(\t)$ is
\eqn\singl{ f(\t)=\sum_{i}a_i t_i\,,}
where each $t_i$ is a letter corresponding to a fundamental
field
and $a_i$ are signs, being $+1$ for a bosonic field, or $-1$
for a fermionic field.

For compact gauge Lie group $G$, the multi-trace partition function is then
given by \ska, (see \refs{\sund,\raam} for refinements,) \eqn\mtrace{
\Z_{G}(\t)=\int_{G}\d \mu_G(g) \exp\Big(\sum_{n=1}^\infty {1\over n}
f(\t^n)\chi_{R}(g^n)\Big) \, , } involving the Haar (or $G$-invariant, or
Hurwitz) measure $\d \mu_G(g)$ for $g\in G$ (so that $\int_G \d\mu_G(g)
F(g)=\int_G \d \mu_G(g) F(g h)=\int_G \d \mu_G(g) F(h g)$ for all $h\in G$
and $\int_G \d \mu_G(g)=1$) and where $\chi_{R}(g)$ is the character for the
$R$ representation of $G$, assuming that the fundamental fields transform in
identical gauge group representations $R$.

For $G=U(N)$, so that for any matrix $U\in U(N)$ we may write $U=V \Theta
V^{\dagger}$, where $V$ is a unitary matrix and $\Theta={\rm
diag}.(e^{i\theta_1},\dots, e^{i \theta_N})$, $0\leq \theta_i < 2\pi$, and
for some $F(U)=F(\Theta)$, independent of $V$, then we may write
\eqn\measures{ \int_{U(N)}\d \mu_{U(N)}(U)F\left(U \right)={1\over (2\pi)^N
N!}\int_0^{2\pi}\prod_{j=1}^N\d \theta_j \prod_{1\leq k< l\leq
N}|e^{i\theta_k}-e^{i\theta_l}|^2\, F(\Theta) \, , } which is, of course,
related to the Weyl parametrisation of $U(N)$. Thus, for such $F(U)$, the
left-hand side of \measures\ simplifies to an integral over the $N$ torus.
Of course, as $\chi_R(U)$ generally depends on linear combinations of
$\tr(U^j)^k\tr(U^{\dagger}{}^l)^m$ for various non-negative integers
$j,k,l,m$ then any function of $\chi_R(U)$ is an example of such an $F(U)$.

We are interested in the case where $R={\rm Adj.}$ is the adjoint
representation so that for $U(N)$ we have that $\chi_{\rm Adj.}(U)=\tr \, U
\, \tr \, U^{\dagger}$ (while for $SU(N)$ then $\chi_{\rm Adj.}(U)=\tr \, U
\, \tr \, U^{\dagger}-1$). For $U(N)$ we then find that, using \mtrace\ with
\measures, \eqn\inttor{ \Z_{U(N)}(\t)={1\over (2\pi)^N
N!}\int_0^{2\pi}\prod_{j=1}^N\d \theta_j \prod_{1\leq k< l\leq
N}|e^{i\theta_k}-e^{i\theta_l}|^2\, \exp\Big(\sum_{n=1}^\infty{1\over
n}f(\t^n)\sum_{j,k=1}^N e^{in(\theta_j-\theta_k)}\Big) \, . }

We may write \inttor\ as an $N$-fold contour integral by first making the
variable change $z_j=e^{i\theta_j}$ so that the integrals in \inttor\ are
around unit circles in each $z_j$ complex plane and then we obtain
\eqn\complextor{ \Z_{U(N)}(\t)={1\over (2\pi i)^N N!}\oint \prod_{i=1}^N {\d
z_i\over z_i}\Delta(\z)\Delta(\z^{-1}) \exp\Big(\sum_{n=1}^\infty{1\over
n}f(\t^n)p_n(\z)p_n(\z^{-1})\Big) \, , } where $\Delta(\z)=\prod_{1\leq
i<j\leq N}(z_i-z_j)$ is the Vandermonde determinant and
$p_n(\z)=\sum_{i=1}^Nz_i{}^n$ is a power symmetric polynomial - see appendix A
for a brief discussion of symmetric polynomials. This integral may then in
principle be evaluated by deforming the contours so as to extract the
residues at poles within the discs $|z_j|<1$, $1\leq j\leq N$.

A crucial observation is that, for some $N$ variable symmetric polynomials
$g(\z),h(\z)$, then \eqn\innero{ \l g,h \r_N=\l h, g\r_N={1\over (2\pi i)^N N!}\oint
\prod_{i=1}^N {\d z_i\over z_i}\Delta(\z) \Delta(\z^{-1}) g(\z)h(\z^{-1}) \,
, } acts as an inner product - this is easy to see in terms of Schur
polynomials which provide an orthonormal basis for symmetric polynomials.
The reader may now wish to peruse appendix A where notation regarding
partitions and a short discussion of symmetric polynomials is included.

\noindent
{\bf The General and Large  $N$ Cases for $U(N)$}

For application of inner products to \complextor\ we have that, in terms of
power symmetric polynomials $p_\lambda(\z)$ for partitions $\lambda$,
\eqn\rewriteo{\eqalign{ \exp\Big(\sum_{n=1}^\infty{1\over
n}f(\t^n)p_n(\z)p_n(\z^{-1})\Big) & =\prod_{n=1}^\infty \sum_{a_n=0}^\infty
{1\over n^{a_n}a_n!} f(\t^n)^{a_n} p_n(\z)^{a_n} p_n(\z^{-1})^{a_n} \, ,\cr
& = \sum_{\lambda} {1\over z_\lambda} f_{\lambda}(\t) p_\lambda(\z)
p_\lambda(\z^{-1}) \, , }} with the definitions of \eqn\newf{
z_\lambda=\prod_{n=1}^\infty n^{a_n} a_n! \, ,\qquad
f_\lambda(\t)=\prod_{n=1}^\infty f(\t^n)^{a_n} \, , } being in terms of the
frequency representation of $\lambda$, $(1^{a_1},2^{a_2},\dots)$ with
$\sum_{n\geq 1} n\, a_n=|\lambda|$, the weight of the partition $\lambda$ (note that
the frequency representation of $\lambda$ is simply a convenient re-ordering
of the parts of $\lambda$).

In \rewriteo\ the numbers $z_\lambda$ have a standard combinatorial
interpretation - for a given permutation $\si\in \S_m$ with $a_1$ 1-cycles,
$a_2$ 2-cycles etc., so that $\sum_{n \geq 1} n\, a_n=m=|\lambda|$, then
$z_\lambda=\prod_{n\geq 1} n^{a_n} a_n!$ is the size of the centraliser
$Z_\si=\{\tau; \tau\in \S_m, \, \tau \si \tau^{-1}=\si\}$ of $\si\in 
\S_m$.  (This
may be easily seen as under conjugation of $\si$ by $\tau$ then $\tau$ can
permute the cycles of length $n$ among themselves in $a_n!$ ways and/or
render a cyclic rotation on each of the individual cycles in $n^{a_n}$
ways.) More details of the symmetric group are to be found in appendix A.

We may immediately observe that \rewriteo\ represents a sum over cycle
polynomials of the symmetric group $\S_m$. This is given by, for letters
$u_1, u_2,\dots u_m$,{\foot{This formula is easy to see from the definition
of the cycle polynomial for a subgroup $G$ of $\S_m$. This is given by $$
{1\over |G|}\sum_{g\in G}u_1{}^{j_1(g)}\cdots u_m{}^{j_m(g)}={1\over 
|G|}\sum_{K_g}|K_g|u_1{}^{j_1(g)}\cdots u_m{}^{j_m(g)} \, , $$ where 
$j_i(g)$ denotes the number of $i$ cycles in the unique decomposition of $g$
into disjoint cycles and $K_g$ denotes the conjugacy classes of $G$ with
class representatives $g$. The size of the conjugacy class $K_g$ is given by
$|K_g|=|G|/|Z_g|$ where $Z_g$ is the centraliser of $g\in G$. For the
present case
then, $G=\S_m$, $|\S_m|=m!$, and $|Z_\si|=z_\lambda$, where $\lambda$ 
gives the cycle structure of $\si\in \S_m$, and thus, for the corresponding
conjugacy class $K_\lambda$, $|K_\lambda|=m!/z_\lambda$.}} \eqn\cyclepoly{
C_m(\u)=\sum_{a_1,\dots,a_m\geq 0}\de_{a_1+2a_2+\dots+m
a_m,m}\prod_{n=1}^m{1\over n^{a_n} a_n!} \, u_n{}^{a_n}= \sum_{\lambda
\vdash m}{1\over z_\lambda} u_{\lambda} \, , } where `$\lambda\vdash m$'
means that $\lambda$ is any partition of $m$ - see appendix A for notation -
and \eqn\ulam{ u_\lambda=\prod_{n=1}^m u_n{}^{a_n} \, , } in terms of the
frequency representation of $\lambda$ above. Identifying
$u_n=f(\t^n)p_n(\z)p_n(\z^{-1})$ then we may rewrite \eqn\rewitecyc{
\Z_{U(N)}(\t)=\sum_{m=0}^\infty {1\over (2\pi i)^N N!}\oint \prod_{i=1}^N
{\d z_i\over z_i}\Delta(\z)\Delta(\z^{-1})
C_m(\u) \, , } the sum of the $U(N)$ group averages of each of the cycle 
polynomials $C_m(\u)$. (Physically, the interpretation is that the cycle
index for the symmetric permutation group accounts for particle statistics
while integration over the gauge group imposes the gauge singlet condition.
For purposes of clarity, the $U(N)$ case has been focused upon here, though
from the form of \mtrace\ it is easy to see how this generalises for
other gauge groups whereby the letters $u_n=f(\t^n)\chi_R(g^n)$
for the fundamental fields transforming in identical gauge group representations, $R$.)

Directly from \rewriteo, in terms of the inner product \innero, then \eqn\partN{
\Z_{U(N)}(\t)=\sum_{\lambda} {1\over z_\lambda} f_{\lambda}(\t) \l
p_\lambda, p_\lambda \r_N =\sum_{\lambda}{1\over
z_{\lambda}}\,f_{\lambda}(\t) \, \sum_{\mu\vdash |\lambda| \atop
\ell(\mu)\leq N}\big(\chi^{\mu}_\lambda\big)^2 \, , } where on the
right-hand side of \partN\ we have used an expression for the inner product
of two power symmetric polynomials expressed in terms of the characters of
the symmetric group, given in appendix A. (Here `$\ell(\mu)$' means the
number of non-zero parts of the partition $\mu$.)

Using a result of appendix A (essentially orthogonality relations for symmetric group
characters), \partN\ may be rewritten as \eqn\approxpartN{
\Z_{U(N)}(\t)=\sum_{\lambda\atop|\lambda|\leq N}f_\lambda(\t)
+\sum_{\lambda\atop |\lambda|>N}{1\over z_{\lambda}}\,f_{\lambda}(\t) \,
\sum_{\mu\vdash |\lambda| \atop \ell(\mu)\leq
N}\big(\chi^{\mu}_\lambda\big)^2 \, . }

In the large $N$ limit, $\Z_{U(N)}(\t)$ simplifies considerably as only
the first term in \approxpartN\ need be considered. Using the frequency
representation of $\lambda$ then \eqn\infNZ{
\Z_{U(\infty)}(\t)=\sum_{\lambda}f_\lambda(\t)= \prod_{n=1}^\infty
\sum_{a_n=0}^\infty f(\t^n)^{a_n}=\prod_{n=1}^\infty{1\over 1- f(\t^n)} \, ,
} a result which has been obtained using Polya counting methods for single
trace operators and saddle point approximations \refs{\sund,\raam}.

Higher order corrections in $|\lambda|$, the weight of the partition
$\lambda$, to \approxpartN\ may be obtained by successive evaluation of
$\sum_{\mu\vdash |\lambda| \atop \ell(\mu)\leq
N}\big(\chi^{\mu}_\lambda\big)^2$. One method is to employ the
Murnaghan-Nakayama Rule, used to compute $\chi^\mu_\lambda$ using skew hooks
and Young diagrams. (A readable account of the Murnaghan-Nakayama Rule may
be found in \sag, though of course it is explained in many standard
textbooks that discuss the symmetric group.)

For the case of $|\lambda|=N{+1}$ then we may observe that, \eqn\caseMNone{
\sum_{\mu\vdash N+1 \atop \ell(\mu)\leq
N}\big(\chi^{\mu}_\lambda\big)^2=\sum_{\mu\vdash N+1 \atop \ell(\mu)\leq
N+1}\big(\chi^{\mu}_\lambda\big)^2- (\chi^{\nu}_\lambda)^2
=z_\lambda-(\chi^{\nu}_\lambda)^2\,,\qquad \nu=(1^{N{+1}}) \, , } since the
partition $\nu=(1^{N+1})$ is the only one excluded among those partitions
$\mu$ of $N{+1}$ with $\ell(\mu)\leq N$. By applying the Murnaghan-Nakayama
Rule we may determine, for $\nu=(1^{L})$, \eqn\mnakao{
\left(\chi^{\nu}_\lambda\right)^2=\cases{1 & for $|\lambda|= L$ \cr 0 &
otherwise} \, , } since $\chi^{\nu}_\lambda$ in this case is just a sign.
(This may be easily seen as there is only one possible way to remove
successive skew hooks, which in this case are just column Young diagrams of
length $\lambda_i$, from the $(1^{L})$ column Young diagram to leave one of
normal shape, in this case, another column Young diagram.) Thus, using
\caseMNone\ with \mnakao\ in \approxpartN, we obtain \eqn\secapproxpartN{
\Z_{U(N)}(\t)=\sum_{\lambda\atop|\lambda|\leq N{+1}}f_\lambda(\t)
-\sum_{\lambda\atop |\lambda|= N{+1}}{1\over z_{\lambda}}\,f_{\lambda}(\t)
+\sum_{\lambda\atop |\lambda|>{N{+1}}}{1\over z_{\lambda}}\,f_{\lambda}(\t)
\,\sum_{\mu\vdash |\lambda| \atop \ell(\mu)\leq
N}\big(\chi^{\mu}_\lambda\big)^2 \, . }

By a similar line of argument we may do the same for the case of
$|\lambda|=N{+2}$. We have that, for $\nu_1=(1^{N{+2}})$ and
$\nu_2=(2,1^{N{+1}})$, \eqn\caseMNoneo{ \sum_{\mu\vdash N+2 \atop
\ell(\mu)\leq N}\big(\chi^{\mu}_\lambda\big)^2=\sum_{\mu\vdash N+2 \atop
\ell(\mu)\leq N+2}\big(\chi^{\mu}_\lambda\big)^2-
(\chi^{\nu_1}_\lambda)^2-(\chi^{\nu_2}_\lambda)^2
=z_\lambda-(\chi^{\nu_1}_\lambda)^2-(\chi^{\nu_2}_\lambda)^2 \, . } We may
determine, for $\nu=(2,1^L)$, \eqn\mnakaol{
\left(\chi^{\nu}_\lambda\right)^2=\cases{(a_1-1)^2 & for
$|\lambda|=\sum_{n\geq 1}n\,a_n=L{+2}$ \cr 0 & otherwise} \, . } Using
\mnakao, \mnakaol\ with \caseMNoneo\ in \secapproxpartN\ then we obtain
\eqn\secapproxpartN{\eqalign{ \Z_{U(N)}(\t)=&\sum_{\lambda\atop|\lambda|\leq
N{+2}}f_\lambda(\t) -\sum_{\lambda\atop N{+1}\leq |\lambda|\leq
N{+2}}{1\over z_{\lambda}}\,f_{\lambda}(\t)-\sum_{\lambda\atop |\lambda|=
N{+2}}{1\over z'_{\lambda}}\,f_{\lambda}(\t)\cr & +\sum_{\lambda\atop 
|\lambda|>{N{+2}}}{1\over z_{\lambda}}\,f_{\lambda}(\t) \,\sum_{\mu\vdash
|\lambda| \atop \ell(\mu)\leq N}\big(\chi^{\mu}_\lambda\big)^2 \, ,} } 
where, in the frequency representation of $\lambda$, \eqn\zprim{ {1\over
z'_{\lambda}}={(a_1{-1})^{2}\over z_\lambda }=\Big({1\over a_1!}-{1\over
(a_1-1)!}+{1\over (a_1-2)!}\Big)\Big/\prod_{n=2}^\infty n^{a_n} a_n! \, . }

We may proceed in this manner to compute explicit higher order corrections
though this becomes cumbersome save for the first few cases as shown. (For
$|\lambda|> N+2$ the corrections will always involve contributions from
\mnakao\ and \mnakaol\ as well as extra ones coming from $\sum_{\mu\vdash 
|\lambda| \atop \ell(\mu)\leq
|\lambda|}\big(\chi^{\mu}_\lambda\big)^2-\sum_{\mu\vdash |\lambda| \atop 
\ell(\mu)\leq N}\big(\chi^{\mu}_\lambda\big)^2$.)

\noindent
{\bf The One Boson Case for $U(N)$}

For the case of one bosonic fundamental field (applicable to half BPS
operators for $\N=4$ super Yang Mills), we have $f(t)=t$
in \complextor, so that we may write \eqn\complextoroneb{
\Z_{U(N)}(t)={1\over (2\pi i)^N N!}\oint \prod_{i=1}^N {\d z_i\over
z_i}\Delta(\z) \Delta(\z^{-1})\prod_{j,k=1}^N {1\over 1-t z_j z_k{}^{-1}} \,
. } 
To evaluate this integral we may use the Cauchy-Littlewood formula, \eqn\cauchlit{
\prod_{i=1}^L\prod_{j=1}^M{1\over 1-x_i y_j}=\sum_{\lambda\atop
\ell(\lambda)\leq {\rm min.}\{L,M\}}s_{\lambda}(x_1,\dots,x_L)
s_\lambda(y_1,\dots,y_M) \, , } where the sum on the right-hand side is
over all partitions $\lambda$ such that the corresponding Young diagrams
have no more than ${\rm min.}\{L,M\}$ rows, $\ell(\lambda)\leq {\rm
min.}\{L,M\}$. With $x_i=t z_i$, $y_i=z_i{}^{-1}$, $i=1,\dots,N$, in \cauchlit, so
that $s_\lambda(t z_1,\dots,t z_N)=t^{|\lambda|}s_\lambda(\z)$,
and employing also
\innero\ and the orthonormality of Schur polynomials, we may easily obtain,
\eqn\complu{ \Z_{U(N)}(t)=\sum_{\lambda\atop \ell(\lambda)\leq
N}t^{|\lambda|}\l s_{\lambda},s_\lambda\r_N=\sum_{\lambda\atop
\ell(\lambda)\leq N}t^{|\lambda|} \, . } By changing summation variables so that
$\lambda_i-\lambda_{i+1}=a_i, i=1,\dots,N{-1}, \lambda_N=a_N$ then we may
write \eqn\compan{ \Z_{U(N)}(t)=\sum_{a_1, \dots,a_N=0}^\infty
t^{a_1+2a_2+\dots + N a_N}=P_N(t) \, , }
where{\foot{$1/P_\infty(t)=\prod_{n=1}^\infty(1-t^n)$ is commonly called the
Euler function, denoted by $\Phi(t)$. $P_\infty(t)=\sum_{n=0}^\infty
p(n)t^n$ acts as a generating function for the number of unordered
partitions of $n$, $p(n)$. Note that $p_N(n)=p(n)$ for $n\leq N$, i.e. the
number of partitions of $n$ into no more than $N$ parts is the same as the
total number of partitions of $n$ so long as $n\leq N$.}} \eqn\defpnt{
P_N(t)=\prod_{i=1}^N{1\over 1-t^i} \, . } Of course this is nothing other
than the generating function for the number $p_N(n)$ of partitions of $n$
into no more than $N$ parts since,
by definition, \eqn\sinces{ \sum_{\lambda\atop
\ell(\lambda)\leq N}\de_{|\lambda|,n}=p_N(n) \, , } so that by the
above
 \eqn\klpe{
\sum_{n=0}^\infty p_N(n) t^n=P_N(t) \, . } This makes explicit the
connection between $\Z_{U(N)}(t)$ and the partition numbers $p_N(n)$.

\noindent
{\bf The  $SU(2)$ Gauge Group Case}

Here we first consider $f(\t)=\sum_{j=1}^kt_j$ in \singl\ so that the
variables $0\leq t_i<1$ represent $k$ bosons in the single particle
partition function. For such fields transforming in the adjoint representation of
$SU(2)$ then \mtrace\ simplifies significantly.  For any $U\in SU(2)$ we may write
$U=V\Theta V^{\dagger}$, where $V$ is unitary and $\Theta={\rm diag.}(e^{i\theta},e^{-i\theta})$,
for $0\leq \theta< 2\pi$,
so that for $F(U)=F(\theta)$, then in usual Weyl parametrisation,
\eqn\sutwoint{\eqalign{
\int_{SU(2)}\d\mu_{SU(2)}(U)F(U)&{}={1\over \pi}\int_0^{2\pi}\d \theta \, \sin^2\theta \,\, F(\theta) \cr
&{}={1\over 4 \pi}\int_0^{4\pi}\d \theta \, \big(1-\cos \theta\big) \, F({\ts{\theta\over 2}})\cr
&{}={1\over 2 \pi}\int_0^{2\pi} \d \theta \,
\big(1-\cos \theta\big)\,F({\ts {\theta\over 2}})\,,}
}
where $F(\theta)=F(\theta+\pi)$ is assumed in writing the last line.
In the present case, $F(U)=F(\theta)=\sum_{n\geq 1}f(\t^n)\chi_{\rm Adj.}(U^n)/n$, where $\chi_{\rm Adj.}(U)=
\tr(U)\tr(U^\dagger)-1=e^{2i\theta}+e^{-2i\theta}+1=2\cos 2\theta +1$, so that
\eqn\sutwoparto{
\Z_{SU(2)}(\t)={1\over 2 \pi}\int_0^{2\pi} \d \theta \, (1-\cos \theta) \prod_{j=1}^k{1\over
(1-t_j)(1-t_j e^{i\theta})(1-t_j e^{-i\theta})}\,.
}
Making the variable change $z=e^{i\theta}$, and using that $F(\theta)=F(-\theta)$ is even, then
\eqn\sutw{
\Z_{SU(2)}(\t)={1\over 2\pi i}\oint {\d z\over z}(1-z)\prod_{j=1}^k{1\over
(1-t_j)(1-t_j z)(1-t_j z^{-1})} \, , }
where the integral is around the unit circle $|z|=1$. 
The residues in \sutw\ may be easily computed
since all the relevant (simple) poles in the disc $|z|<1$ occur at the
points $z=t_j$. Thus \eqn\ergyt{
\Z_{SU(2)}(\t)=\sum_{i=1}^k{t_i{}^{k-1}\over 1-t_i{}^2}\prod_{j=1\atop j\neq
i}^k{1\over (t_i-t_j)(1-t_it_j)(1-t_j)} \, . }

This partition function has an interesting interpretation
from a group theory perspective. We may write,
 \eqn\gyu{ {1+t\over
(1-t z)(1-t z^{-1})}=\sum_{n=0}^\infty \chi_{n}(z) t^n \, , } where
\eqn\spiutc{ \chi_j(z)={z^{j+{1\over 2}}-z^{-j-{1\over 2}}\over z^{1\over 2}-z^{-{1\over 2}}} 
\, ,\qquad j\in \half \Bbb{Z}\,, } is an $SU(2)$
character, corresponding to the spin $j$ irreducible representation,
$R_{j}$. Now the integral in \sutw\ acts as an $SU(2)$ inner
product, \eqn\innersu{ \l \chi_j,\chi_k\r={1\over 2 \pi i}\oint {\d z\over
z} (1-z) \chi_j(z) \chi_k(z^{-1})=\de_{jk} \,, } for $j,k\in \Bbb{N}$.
Thus, from \sutw\ with \gyu\ and \innersu,
\eqn\frt{
\prod_{j=1}^k(1-t_j{}^2)\Z_{SU(2)}(\t)=\sum_{n_1,\dots,n_k=0}^\infty \l
\chi_{n_1}\cdots \chi_{n_k},1\r t_1{}^{n_1}\cdots t_k{}^{n_k} \, , } acts
as a generating function for the number of singlets in the decomposition of
the $SU(2)$ representation $R_{n_1}\otimes \cdots \otimes R_{n_k}$.{\foot{
Generating functions for products of Lie algebra representations have been
considered elsewhere, in \cumm\ for instance. A generating function for the
number of singlets in $n$ products of the fundamental times $n$ products of
the anti-fundamental representations for $SU(N)$ was found by Gessel \gess\
in terms of Toeplitz determinants involving Bessel functions. See also
\carl\ for a nice physics oriented discussion of similar issues. The special
case of $R_{1\over 2}\otimes \cdots \otimes R_{1\over 2}$ ($2n$ products of
the fundamental) for $SU(2)$, contains a Catalan number, ${1\over n+1}\left({
2n\atop n}\right)$, of singlet representations.}} By using that
$\chi_{n}(z)=\sum_{j=-n}^nz^{j}$ we may use the Cauchy residue theorem to
compute explicitly that \eqn\cauh{ \l \chi_{n_1}\cdots
\chi_{n_k},1\r=\sum_{j_1=0}^{2n_1}\cdots
\sum_{j_k=0}^{2n_k}\left(\de_{j_1+\cdots +j_k, n_1+\cdots+ n_k}-
\de_{j_1+\cdots + j_k,n_1+\cdots + n_k+1}\right) \,. }

If we modify the one particle partition function to include $k$ bosons and
${\bar k}$ fermions and hence consider \singl\ in the form 
$f(t,{\bar
t})=\sum_{j=1}^k t_j-\sum_{{\bj}=1}^{\bar k}{\bar t}_{\bj}$
then we may similarly as above evaluate \eqn\sutwbosfer{
\Z_{SU(2)}(\t,{\bar \t})= {1\over 2\pi i}\oint {\d z\over
z}(1-z)\prod_{1\leq j\leq k\atop 1\leq \bj \leq {\bar k}}{(1-{\bar
t}_{\bj}) (1-{\bar t}_{\bj} z)(1-{\bar t}_{\bj} z^{-1})\over (1-t_j)(1-t_j
z)(1-t_j z^{-1})}
\, ,}  
where the contour is around the unit disc $|z|=1$.
So long as $k>{\bar k}$ then \sutwbosfer\ receives contributions 
from only those simple poles at $z=t_j$ so that for this case we obtain
\eqn\sutwbosferres{ \Z_{SU(2)}(\t,{\bar \t})\Big|_{k>{\bar
k}}=\sum_{i=1}^k{t_i{}^{k-{\bar k}-1}\over 1-t_i{}^2}\prod_{1\leq \bj \leq
{\bar k}\atop 1\leq j\leq k, j\neq i}{(t_i-{\bar t}_{\bj})(1-t_i{\bar
t}_{\bj})(1-{\bar t}_{\bj})\over (t_i-t_j)(1-t_it_j)(1-t_j)} \, . } For
$k\leq {\bar k}$ then \sutwbosfer\ also receives contributions from poles at
$z=0$. For instance \eqn\sutwbosferres{ \Z_{SU(2)}(\t,{\bar
\t})\Big|_{k={\bar k}}=\sum_{i=1}^k{1\over t_i(1-t_i{}^2)}\prod_{1\leq j,
\bj \leq k\atop j\neq i}{(t_i-{\bar t}_{\bj})(1-t_i{\bar t}_{\bj})(1-{\bar
t}_{\bj})\over (t_i-t_j)(1-t_it_j)(1-t_j)} + \prod_{1\leq j, \bj\leq
k}{{\bar t}_{\bj}(1-{\bar t}_{\bj})\over t_j(1-t_j)}\, , } where the last
term on the right-hand side of \sutwbosferres\ comes from the simple pole at
$z=0$.

These formulae should be useful for computing the multi-trace partition
functions, for fundamental fields transforming in an $SU(2)$ gauge group, in
other sectors of $\N=4$ super Yang Mills. For instance, after a suitable
identification of the variables $t_j,{\bar t}_{\bj}$ with
variables in single particle partition functions for semi-short sectors of
$\N=4$ super Yang Mills, described in detail in \char, then \sutwbosferres\
should allow for an explicit expression for corresponding multi-trace
partition functions. They may also be useful for computing the $\N=4$ 
superconformal index
of \mald\ for $SU(2)$ gauge group, or at least for restrictions of it such
as described in \char\ or \nak.

\noindent
\newsec{Counting Operators in Free $\N=4$ Super Yang Mills}

In this section, the counting of half and quarter BPS
operators for free $\N=4$ super Yang Mills, when the fundamental
fields transform in the adjoint representation of $U(N)$, is discussed
in some detail.

\noindent
{\bf Counting Operators Directly}

We may, of course, proceed to count multi-trace half and quarter BPS
primary 
operators directly, in terms of the
fundamental fields, $Z$, for half BPS operators and $Z,Y$,
for quarter BPS operators.   

$(Z,Y)$ forms a $U(2)$
doublet, where $U(2)$ has generators given by a subset of the $SU(4)_R$
generators, $H_i,E_{i\pm}$, $1\leq i\leq 3$, where $H_i$ are the Cartan
sub-algebra generators and $E_{i\pm}$ are ladder operators satisfying
(in the Chevalley-Serre basis)
$[H_i,E_{j\pm}]=\pm K_{ij}E_{j\pm}$, with $[K_{ij}]$ being the usual
$SU(4)$
Cartan matrix.
The $U(2)$ generators consist of the  $SU(2)$ generators $H_2,E_{2\pm}$, where explicitly
$[(H_1,H_2,H_3),E_{2\pm}]=\mp(1,-2,1)E_{2\pm}$, along with the generator
$H_1{+}H_2{+}H_3$, whose eigenvalues give the conformal dimensions in this
case, $[H_1{+}H_2{+}H_3,(Z,Y)]=(Z,Y)$ \char.  Explicitly, we have that $[E_{2+},Z]=0$,
$[E_{2-},Z]=Y$, $[(H_1,H_2,H_3),Z]=(0,1,0)Z$,
$[(H_1,H_2,H_3),Y]=(1,-1,1)Y$ so that an operator involving $n$ $Z$'s
and $m$ $Y$'s
transforms
in the $[m,n{-m},m]$ $SU(4)_R$ $R$-symmetry representation.

For $k$-trace half
BPS primary operators transforming in the
$[0,n,0]$ $SU(4)_R$ $R$-symmetry representation, with
conformal dimension $n$, then in terms of the fundamental
field $Z$ a basis is provided by,
\eqn\formofhalf{
{\rm tr}(Z^{n_1})\cdots {\rm tr}(Z^{n_k}) \, , \quad \sum_{i=1}^kn_i=n \,
.
}
We have that, due to trace identities for finite $N$, ${\rm tr}(Z^n)$
for
$n>N$ is expressible in terms of a sum over multi-trace operators
of the form \formofhalf, for $k>1$, and thus, a minimal basis for
multi-trace half BPS primary operators consists of \formofhalf\ for
all $1\leq k\leq n$ and with every $n_i\leq N$, ordered so that
$n_1\geq n_2\geq\dots\geq n_k\geq 0$, i.e. so that
$(n_1,\dots,n_k)$
is a partition of $n$ where each part $n_i\leq N$.
With this restriction, the number of multi-trace half BPS primary operators
for a given $n$ is 
\eqn\flambdao{
\N_{(n)}=p_N(n) \, , }
since the number $p_N(n)$, in \sinces, of
partitions
of $n$ into $\leq N$ parts is the same as the number of partitions
of $n$ in which each part is $\leq N$ -
see \andrews\ for a simple proof employing generating functions.

For quarter BPS operators belonging to the $[m,n{-}m,m]$ $SU(4)_R$
$R$-symmetry representation,
a basis for $k$-trace operators is
\eqn\basquar{
{\rm tr}\Big(\prod_j Z^{n_{1j}}Y^{m_{1j}}\Big)\cdots 
{\rm tr}\Big(\prod_j Z^{n_{kj}}Y^{m_{kj}}\Big) \,, \quad
\sum_{i,j}n_{ij}=n\,,
\quad \sum_{i,j}m_{ij}=m \, ,
}
where there is a choice of ordering in each trace.
(Note that the $m=0$
case corresponds to the half BPS case already considered.)
Using the basis provided by \basquar\ for all
allowable $k$, then to avoid over-counting of multi-trace quarter BPS 
 operators, the cyclicity of each trace and also trace identities for
finite $N$ must be accounted for.  Assuming that this is done, let
$\M_{(n,m)}$ denote the number of elements in this minimal basis
for multi-trace operators of the form \basquar.
Then,  to obtain the number $\N_{(n,m)}$ of  multi-trace quarter BPS
{\it primary} operators
in the $SU(4)_R$ representation $[m,n{-}m,m]$, the
number of $U(2)$
descendants, in the $SU(4)_R$ representation $[m,n{-}m,m]$, of
multi-trace quarter 
BPS primary operators, in 
$SU(4)_R$ representations $[j,n{+}m{-}2j,j]$,
$0\leq j\leq m{-1}$, must be subtracted from $\M_{(n,m)}$.  (These
descendants  arise due to
the relation $[E_{2-},Z]=Y$. Acting with $(E_{2-}){}^{m-j}$
on the highest weight state in the $SU(4)_R$ representation
$[j,n{+}m{-}2j,j]$
we obtain a descendant in the $SU(4)_R$ representation $[m,n{-}m,m]$.)
The number of such $U(2)$ descendants coincides with
$\N_{(n{+}m{-}j,j)}$, the number of corresponding primary operators.
In this way, we obtain that
\eqn\wormhead{
\M_{(n,m)}=\N_{(n,m)}+\N_{(n+1,m-1)}+\dots +\N_{(n+m-1,1)}+\N_{(n+m)}
\, ,
}
so that $\N_{(n,m)}=\M_{(n,m)}-\M_{(n+1,m-1)}$ 
may be obtained recursively for each $m$.

We may illustrate by counting all multi-trace quarter BPS
primary operators in the $[1,n{-1},1]$ $R$-symmetry representation.  In this case
a basis for $k{+1}$-trace operators is provided by
\eqn\meqlonec{
{\rm tr}(Z^{n_1})\cdots {\rm tr}(Z^{n_{k}}){\rm tr}(Z^j Y) \, ,
\qquad
\sum_{i=1}^{k}n_i=n-j \, .
}
Cyclicity of  traces implies that we may arrange $Y$ as shown,
to avoid over-counting.
$U(N)$ trace identities imply, similarly as for the half BPS case,
 that a minimal basis for
multi-trace operators requires $j<N$ and each $n_i\leq N$
in \meqlonec\ for every $1\leq k\leq n{-j}$, so that $(n_1,\dots,n_k)$
forms a partition of $n{-j}$, with every part $\leq N$.  Thus, by a similar
argument as for the half BPS case, $\M_{(n,1)}=\sum_{j=0}^{N-1}p_N(n-j)$.
Finally, to
ensure that only primary operators are counted then we
must subtract off contributions from descendants of
half BPS primary operators in the $[0,n{+1},0]$ $SU(4)_R$
representation, of which there are $p_{N}(n{+1})$.
Using \wormhead\ with \flambdao\ we then conclude that
\eqn\countingonen{
\N_{(n,1)}=\sum_{j=0}^{N-1}p_N(n-j)-p_{N}(n{+1}) \, ,
}
gives the number of
multi-trace quarter BPS primary operators in the $[1,n{-1},1]$
$R$-symmetry representation.

Counting in this fashion becomes more difficult for greater $m$ and
now a procedure is described employing symmetric polynomials 
to find a generating function for the numbers of 
multi-trace quarter BPS primary operators in the $[m,n{-}m,m]$
$SU(4)_R$ representation, for $m=0,1,2$ at finite $N$ and
for any $n,m$ at
large $N$.  This generating function is subsequently used to provide asymptotic counting for
fixed $m$, large $n$ in the large $N$ limit.

\noindent
{\bf Counting Operators via Expansion of Partition Functions in
Schur Polynomials}

For $k$ bosonic fundamental fields, we may take
$f(\t)=\sum_{j=1}^kt_j$ in \singl\ so 
that \complextor\ may be
written as \eqn\complextoronebpre{ \Z_{U(N)}(\t)={1\over (2\pi i)^N N!}\oint
\prod_{i=1}^N {\d z_i\over
z_i}\Delta(\z)\Delta(\z^{-1})\prod_{j=1}^k\prod_{r,s=1}^N {1\over 1-t_j z_r
z_s{}^{-1}} \, . } Often it is the case that such partition functions should
be expanded in terms of $s_\lambda(\t)$, the $k$ variable Schur polynomial
labelled by partitions $\lambda$. An example is provided by \exptwobosc\ for
counting multi-trace quarter BPS operators. We may use the Cauchy-Littlewood
formula \cauchlit\ to expand in this way, to obtain \eqn\multibos{
\Z_{U(N)}(\t)=\sum_{\lambda\atop \ell(\lambda)\leq {\rm min.}\{k,N^2\}}
\!\!\! \N_{\lambda} \, s_\lambda(\t) \, , } where \eqn\flam{
\N_\lambda={1\over (2\pi i)^N N!}\oint \prod_{i=1}^N {\d z_i\over
z_i}\Delta(\z)\Delta(\z^{-1})\,s_\lambda(\z \z^{-1}) \, ,}
where $\z \z^{-1}$ has components $z_i z_j{}^{-1}$, $1\leq i,j\leq N$.

{}From Macdonald \mac\ we have that \eqn\knoem{ s_\lambda(\x
\y)=\sum_{\mu,\nu\vdash|\lambda|} \gamma_{\mu\nu}^\lambda s_\mu(\x)s_\nu(\y)
\, , } in terms of Kronecker coefficients, \eqn\syt{\eqalign{
\gamma^{\lambda}_{\mu\nu}=& {1\over 
|\lambda|!}\sum_{\si\in 
\S_{|\lambda|}}\chi^{\lambda}(\si)\chi^{\mu}(\si)\chi^\nu(\si)\cr =&
\sum_{\rho\vdash |\lambda|}{1\over z_\rho} \chi^\lambda_\rho \,\chi^\mu_\rho
\, \chi^\nu_\rho \, ,} } being a sum over irreducible $\S_{|\lambda|}$
characters evaluated at $\si\in\S_{|\lambda|}$, related to a sum over
irreducible $\S_{|\lambda|}$ characters evaluated on the conjugacy classes
labelled by the partitions $\rho$ in the second line. Using \innero\ along
with the orthonormality property of Schur polynomials we find that \eqn\styp{
\N_\lambda=\sum_{\mu\vdash |\lambda|\atop \ell(\mu)\leq
N}\gamma^{\lambda}_{\mu \mu} \, . }

The situation becomes much more involved if we include also $\bar k$
fermionic fields, so that \singl\ may be written in the form $f(t,{\bar
t})=\sum_{j=1}^k t_j-\sum_{{\bj}=1}^{\bar k}{\bar t}_{\bj}$, and attempt to expand $\Z_{U(N)}(\t, {\bar
\t})$ in terms of products of Schur polynomials $s_\lambda(\t) s_\mu({\bar
\t})$. Such expansion is required for counting, for instance, for the free
field partition function in the eighth BPS sector of $\N=4$ super Yang
Mills.  In this case the partition function is expanded, analogous to \exptwobosc, in
terms of $SU(2|3)$ characters, which may be expressed in terms of a
linear combination of products of two-variable and three-variable Schur
polynomials. (See \char\ for a discussion of counting for the eighth BPS sector
along these lines.) Including fermions,
\complextoronebpre\ becomes modified by \eqn\complextoronebref{
\Z_{U(N)}(\t,{\bar \t})={1\over (2\pi i)^N N!}\oint \prod_{i=1}^N {\d
z_i\over z_i}\Delta(\z)\Delta(\z^{-1})\prod_{1\leq j\leq k\atop1\leq \bj
\leq {\bar k}} \prod_{r,s=1}^N {1-{\bar t}_{\bj} z_r z_s{}^{-1}\over 1-t_j
z_r z_s{}^{-1}} \, . } To achieve the expansion, we may use the
Cauchy-Littlewood formula \cauchlit\ along with another formula of
Littlewood, \eqn\littlewood{ \prod_{i=1}^L \prod_{j=1}^M(1+x_i
y_j)=\sum_{\lambda \atop \ell(\lambda)\leq L, \ell({\tilde \lambda})\leq
M}s_\lambda(x_1,\dots,x_L) \, s_{{\tilde \lambda}}(y_1,\dots,y_M) \, , }
where ${\tilde \lambda}$ is the partition conjugate to $\lambda$ (where the
rows and columns of the Young diagram corresponding to $\lambda$ are
interchanged) and where the sum is restricted to those $\lambda$ whereby the
corresponding Young diagrams have at most $L$ rows, $\ell(\lambda)\leq L$,
and $M$ columns, $\ell({\tilde \lambda})\leq M$. We may thus write
\eqn\multibosfer{ \Z_{U(N)}(\t,{\bar \t})=\sum_{\lambda\atop
\ell(\lambda)\leq {\rm min.}\{k,N^2\}} \sum_{\mu\atop \ell(\mu)\leq {\bar
k}, \ell({\tilde \mu})\leq N^2} \!\!\! \N_{\lambda,\mu} \, s_\lambda(\t) \,
s_\mu({\bar \t}) \, , } where \eqn\flambosferm{ \N_{\lambda,\mu}={(-1)^{|\mu|}\over
(2\pi i)^N N!}\oint \prod_{i=1}^N {\d z_i\over
z_i}\Delta(\z)\Delta(\z^{-1})\,s_\lambda(\z \z^{-1}) s_{\tilde \mu}(\z
\z^{-1})\, . } Obviously these numbers are considerably more involved than
those in \flam. We may of course use \knoem\ again to interpret
\flambosferm\ in terms of Kronecker coefficients.

\noindent
{\bf Counting Quarter BPS Operators by Symmetric Polynomial Methods}

The two bosonic fundamental field case is now focused
upon.{\foot{The two boson case leads to an interesting generalisation of an
identity in \willen\ involving Littlewood-Richardson coefficients
$c^{\nu}_{\lambda \mu}$, the coefficients that appear in the decomposition
$s_\lambda(\x)s_\mu(\x)=\sum_{\nu} c^\nu_{\lambda \mu} s_\nu(\x)$. With
$f(\t)$ in \singl\ given by $f(\t)=t_1+t_2$, and expanding appropriately the corresponding integrand in
\complextoronebpre\ using \cauchlit; then using \innero, the orthonormality
of Schur polynomials and the result \infNZ, we obtain (note that
$c^\nu_{\lambda \mu}=0$ if $|\nu|\neq |\lambda|+|\mu|$) $$
\Z_{U(\infty)}(t_1,t_2)=\sum_{\lambda,\mu}t_1{}^{|\lambda|}t_2{}^{|\mu|}
\langle s_\lambda s_\mu,s_\lambda s_\mu\rangle_\infty=
\sum_{\lambda,\mu,\nu\atop \nu\vdash 
|\lambda|+|\mu|}t_1{}^{|\lambda|}t_2{}^{|\mu|}(c^{\nu}_{\lambda\mu})^2=
\prod_{n\geq 
1}{1\over 1-t_1{}^n-t_2{}^n} $$ which reduces to Theorem 4.1 of \willen\ if
we take $t_1=t_2=t$.}}  In particular, the numbers $\N_{(n,m)}$
in  \exptwobosc\ are evaluated using results of the last sub-section.

We may proceed to evaluate $\N_\lambda$ recursively. The simplest case is
for $\N_{\lambda}=\N_{(n)}$, whereby introducing a formal variable $t$ then
it is clear, by \cauchlit\ with \complextoroneb, \compan\ and \defpnt, that
\eqn\clarth{ \sum_{n=0}^\infty \N_{(n)} t^n={1\over (2\pi i)^N N!}\oint
\prod_{i=1}^N {\d z_i\over z_i}\Delta(\z) \Delta(\z^{-1})\prod_{j,k=1}^N
{1\over 1-t z_j z_k{}^{-1}}=P_N(t) \,, } so that, by \klpe, $\N_{(n)}$
is
given by
\flambdao.

More generally to evaluate $\N_{(n,m)}$ from \flam\ we may use, for $\y=\z
\z^{-1}$, \eqn\moregen{\eqalign{ s_{(m)}(\y)s_{(n)}(\y) & {} =
s_{(n,m)}(\y)+s_{(n+1,m-1)}(\y)+\dots+s_{(n+m-1,1)}(\y)+s_{(n+m)}(\y) \,
,\cr s_{(m)}(\z\z^{-1})& {}= \sum_{\mu\vdash m\atop \ell(\mu)\leq
N}s_\mu(\z)s_{\mu}(\z^{-1}) \, ,} } 
where the expression in the first line
of \moregen\ may be easily seen using Young tableaux multiplication rules
while \cauchlit\ determines the expression in the second line. 
From \flam\ with \moregen, we may find a
useful generating function, in terms of a formal variable $t$, for the numbers in \wormhead\ as follows,
\eqn\gengre{\eqalign{
\F^{(m)}_N(t)&=\sum_{n=0}^\infty \M_{(n,m)}t^n\cr &={1\over (2\pi i)^N N!}\oint \prod_{i=1}^N
{\d z_i\over z_i}\Delta(\z)\Delta(\z^{-1})\,s_{(m)}(\z \z^{-1})\,
\sum_{n=0}^\infty s_{(n)}(\z \z^{-1}) t^n\cr &={1\over (2\pi i)^N N!}\oint
\prod_{i=1}^N {\d z_i\over z_i}\Delta(\z)\Delta(\z^{-1})\sum_{\mu\vdash
m\atop \ell(\mu)\leq N}s_{\mu}(\z) s_{\mu}(\z^{-1})\prod_{j,k=1}^N {1\over
1-t z_j z_k{}^{-1}}\cr &=\sum_{\mu\vdash m\atop \ell(\mu)\leq
N}\sum_{\lambda}t^{|\lambda|}\l s_{\lambda}s_\mu, s_{\lambda}s_\mu\r_N \, ,}
} so that we may write \eqn\recursNnm{ \N_{(n,m)}={1\over 2 \pi i}\oint {\d
t\over t} \Big({1\over t^{n}}\F^{(m)}_N(t)-{1\over t^{n+1}}
\F^{(m-1)}_N(t)\Big) \, , } which allows for recursive determination of
$\N_{(n,m)}$.

Applying this to the case of $\N_\lambda=\N_{(n,1)}$ we have, from \moregen\
\eqn\notuce{ s_{(1)}(\z \z^{-1})= s_{(1)}(\z)s_{(1)}(\z^{-1}) \, , } so
that, from \gengre, \eqn\extraclarif{
\F^{(1)}_N(t)=\sum_{n=0}^\infty\big(\N_{(n,1)}+\N_{(n+1)}\big)t^n=\sum_{
\lambda\atop \ell(\lambda)\leq N}t^{|\lambda|}\l s_{(1)}
s_\lambda,s_{(1)}s_\lambda\r_N \,. } Using (again, this may be easily seen
from Young tableaux multiplication rules) \eqn\prodty{
s_{(1)}(\z)s_{\lambda}(\z)=\sum_{r=1}^Ns_{\lambda+\e_r}(\z) \, , } for
$\{\e_r; 1\leq r\leq N, \e_r \cdot \e_s=\de_{rs}\}$ being usual orthonormal
vectors, we find that \eqn\greou{\eqalign{
\F^{(1)}_N(t)=\sum_{n=0}^\infty\big(\N_{(n,1)}+\N_{(n+1)}\big)t^n=\sum_{
\lambda\atop \ell(\lambda)\leq N}t^{|\lambda|}\sum_{r,s=1}^N \l
s_{\lambda+\e_r},s_{\lambda+\e_s}\r_N \, . }} Now for any partition
$\lambda$, $\l s_{\lambda+\e_r},s_{\lambda+\e_s}\r_N$ vanishes unless
$\e_r=\e_s$ for any $r,s$ and $\lambda_{r{-1}}-\lambda_{r}>0$ for
$r=2,\dots,N$, due to \eqn\idsf{
s_{(\lambda_1,\dots,\lambda_{r-1},\lambda_{r}+1,\dots,\lambda_{N})}(\z)=0
\quad {\rm for}\quad \lambda_{r-1}=\lambda_{r}\, ,\quad r>1 \, . } Changing
summation variables to those in \compan\ then we have, with the definition
\defpnt, \eqn\complin{\eqalign{ \F^{(1)}_N(t)&
=\sum_{n=0}^\infty\big(\N_{(n,1)}+\N_{(n+1)}\big)t^n
=\sum_{a_1,\dots,a_N\geq 0}t^{a_1+\dots + N
a_N}+\sum_{r=1}^{N-1}\sum_{a_1,\dots,a_N\geq 0\atop a_{r}\geq 1}t^{a_1+\dots
+ N a_N}\cr &= \sum_{i=0}^{N-1} t^i\sum_{a_1,\dots,a_N\geq 0}t^{a_1+\dots +
N a_N}={1\over 1-t}P_{N-1}(t) \, .} } Thus, using \klpe, \flambdao\ with
\complin,{\foot{This formula agrees with \countingonen\ due to
$\sum_{j=0}^{N-1}p_N(n-j)=\sum_{j=0}^np_{N-1}(j)$ which follows
because
the corresponding generating functions match,
$$
\sum_{n=0}^\infty\sum_{j=0}^{N-1}p_N(n-j)t^n=(1+t+\dots+t^{N-1})P_N(t)= {1\over 1-t}P_{N-1}(t)
=\sum_{n=0}^\infty\sum_{j=0}^np_{N-1}(j)t^n
 \, .
$$}}
 \eqn\styo{
\N_{(n,1)}=\sum_{j=0}^np_{N-1}(j)-\N_{(n+1)}=\sum_{j=0}^{n}p_{N-1}(j)-p_N(n+
1) \, . }

{}For the case of $\N_\lambda=\N_{(n,2)}$ we have that, from \moregen,
\eqn\simpk{ s_{(2)}(\z
\z^{-1})=s_{(2)}(\z)s_{(2)}(\z^{-1})+s_{(1,1)}(\z)s_{(1,1)}(\z^{-1}) \, , }
so that, from \gengre, we have \eqn\compiu{ \F^{(2)}_N(t)=\sum_{\lambda\atop
\ell(\lambda)\leq N}t^{|\lambda|} \left(\l
s_{(2)}s_\lambda,s_{(2)}s_\lambda\r_N+\l
s_{(1,1)}s_\lambda,s_{(1,1)}s_\lambda\r_N\right) \, . } Using
\eqn\produy{\eqalign{ s_{(2)}(\z)s_{\lambda}(\z)&
=\sum_{r=1}^Ns_{\lambda+2\e_r}(\z)+\sum_{1\leq r<s\leq
N}s_{\lambda+\e_r+\e_s}(\z) \, ,\cr s_{(1,1)}(\z)s_{\lambda}(\z)& =
\sum_{1\leq r<s\leq N}s_{\lambda+\e_r+\e_s}(\z) \, ,} } along with \idsf\
and \eqn\idfg{ s_{(\lambda_1, \dots,
\lambda_{r-1},\lambda_{r}+2,\dots,\lambda_N)}(\z)=-s_{(\lambda_1,\dots,
\lambda_{r}+1,\lambda_{r-1}+1,\dots,\lambda_N)}(\z) \, , } for the cases where
$\lambda_{r-1}=\lambda_{r}$, we may obtain, with the definition \defpnt,
\eqn\compiuo{ \F^{(2)}_N(t)= {1{-t^{N+1}}\over (1{-t})(1{-t^2})}
P_{N{-1}}(t)+{1\over (1{-t})(1{-t^2})}P_{N{-2}}(t)\, , } where the first
contribution comes from $\sum_{\lambda} t^{|\lambda|} \l
s_{(2)}s_\lambda,s_{(2)}s_\lambda\r_N$ while the second comes from
$\sum_{\lambda} t^{|\lambda|} \l s_{(1,1)}s_\lambda,s_{(1,1)}s_\lambda\r_N$.
Since the partition number $p_k(-n)=0$ for $n=1,2,\dots$ we may write, using
\klpe,{\foot{This is a special case of the following: for any $f(n)$, $n\in
\Bbb{Z}$, that satisfies $f(-n)=0$, $n=1,2,\dots$, then we may (at least
formally) write $$ P_k(t)\sum_{n=0}^\infty
f(n)t^n=\sum_{n,i_1,\dots,i_k=0}^\infty f(n-i_1-2 i_2-\dots-k i_k) t^n \, .
$$}} \eqn\oddth{ {1\over (1-t)(1-t^2)}P_k(t)=\sum_{n,i,j=0}^\infty
p_k(n-i-2j) t^n \, . } Thus, from \compiuo\ with \styo,
\eqn\ndetermined{\eqalign{ &\N_{(n,2)}=-\sum_{j=0}^{n+1} p_{N-1}(j)\cr &
+\cases{\sum_{i,j=0}^\infty
\left(p_{N{-2}}(n{-i}{-2j})+p_{N{-1}}(n{-i}{-2j})\right) & if $n\leq N$,\cr
\sum_{i,j=0}^\infty
\left(p_{N{-2}}(n{-i}{-2j})+p_{N{-1}}(n{-i}{-2j})-p_{N{-1}}(N{+1}{-n}{-i}{-2
j})\right) & if $n\geq N{+1}$.}} }

Tables of the numbers \flambdao, \styo\ and \ndetermined\ are given in
appendix B for some few cases of $n, N$.  Notice from these
tables that the numbers $\N_{(n,m)}$ below the diagonal line
$N\geq n+m$ for a given $n$ are the same for all $N$.  This is
a general feature that derives from values of $\N_{(n,m)}$
for $N\geq n+m$, which numbers may be obtained from a corresponding
generating function that is now constructed.

Using these techniques, we may provide a consistency check of \clarth, \complin, \compiuo\ along
with a general result for $\N_{(n,m)}$ for high enough values of $N$, $N\geq
m+n$. This employs the orthogonality property of power symmetric
polynomials $p_\lambda(\z)$
(in the large $N$ limit) along with \eqn\kropet{\eqalign{ s_{(n)}(\z) &
=\sum_{\lambda\vdash n} {1\over z_\lambda}p_\lambda(\z)\cr &
=\sum_{i_1,\dots,i_n=0}^\infty {1\over i_1!i_2!\cdots
i_n!}\,\de_{i_1{+2}i_2+\dots +ni_n,n}\,p_1(\z)^{i_1}\, \big({\ts{1\over
2}}p_2(\z)\big)^{i_2}\cdots \big({\ts{1\over n}}p_n(\z)\big)^{i_n} \, .} }
Using the trivial identity $p_\lambda(\x \y)=p_\lambda(\x)p_\lambda(\y)$
then from \innero, \gengre\ with \kropet\ we have \eqn\gengrelp{\eqalign{
\F^{(m)}_\infty(t) &=\sum_{n=0}^{\infty}\sum_{\lambda\vdash
n}\sum_{\mu\vdash m}{1\over z_\lambda z_\mu}\l p_{\lambda}p_\mu,p_\lambda
p_\mu\r_\infty \, t^n\cr &=\sum_{n=0}^{\infty}\sum_{\lambda\vdash
n}\sum_{\mu\vdash m} {1\over z_\lambda z_\mu} \l p_\nu,p_\nu\r_\infty \,
t^n\cr &=\sum_{n=0}^{\infty}\sum_{\lambda\vdash n}\sum_{\mu\vdash m}
{z_\nu\over z_\lambda z_\mu} t^n\, ,} } where for $(1^{a_1},2^{a_2},\dots)$
being the frequency representation of $\lambda$ and
$(1^{b_1},2^{b_2},\dots)$ being that of $\mu$ then $\nu$ has frequency
representation $(1^{a_1+a_2},2^{a_2+b_2},\dots)$ so that $|\nu|=n+m$. This
agrees with $\F^{(m)}_N(t)$ in a series expansion up to $O(t^{N-m})$ (since
the last equation in \gengrelp\ is also valid for finite $N$ so long as
$|\nu|=n+m\leq N$, by a result of appendix A). Now, since \eqn\sinceagain{
{z_\nu\over z_\lambda z_\mu}=\prod_{j=1}^\infty {(a_j{+}b_j)!\over a_j!
b_j!} \, , } we obtain from \gengrelp\ that, \eqn\gengrelpp{\eqalign{
\F_\infty^{(m)}(t)&=\sum_{n=0}^\infty\sum_{a_1,\dots,a_n=0}^\infty\sum_{b_1,
\dots,b_m=0}^\infty \de_{a_1+\cdots + n a_n,n}\de_{b_1+\cdots + m b_m,m}
\prod_{j=1}^m{(a_j+b_j)!\over a_j! b_j!} t^n\cr & =
\sum_{b_1,\dots,b_m=0}^\infty\de_{b_1+\cdots + m
b_m,m}\prod_{j=1}^m\sum_{a_j=0}^\infty {(a_j+b_j)!\over a_j! b_j!} t^{j
a_j}\prod_{j>m}{1\over 1-t^j}\cr & =\sum_{b_1,\dots,b_m=0}^\infty
\de_{b_1+\cdots + m b_m,m}\prod_{j=1}^m{1\over
(1-t^j)^{b_j+1}}\prod_{j>m}{1\over 1-t^j} \,. }} For the first few cases we
have that, with $P_N(t)$ as defined in \defpnt, \eqn\refusesack{
\F_\infty^{(m)}(t)=\cases{ P_\infty (t) & for $m=0$ \cr {1\over 1-t}P_\infty
(t) & for $m=1$ \cr {2\over (1-t)(1-t^2)}P_\infty(t) & for $m=2$} \, , }
whose series expansion agrees with \clarth, \complin, \compiuo\ up to
$O(t^{N-m})$ for, respectively, $m=0,1,2$. We may use \recursNnm\ with \gengrelp\ to determine
$\N_{(n,m)}$ exactly for $N\geq n+m$.

\noindent
{\bf Asymptotic Counting of Quarter BPS Operators  at Large $N$}

Asymptotic counting for the one boson case in the large $N$ limit, for
which, with $P_N(t)$ as defined in \defpnt, with $p(n)$ being the total
number of (unordered) partitions of $n$, \eqn\asympob{
\Z_{U(\infty)}(t)=P_\infty(t)=\sum_{n=1}^\infty p(n) t^n \, , } is the
multi-trace partition function, entails finding an asymptotic value for the
partition number $p(n)$ for `large' $n$. This may be achieved by performing
a saddle point approximation of $p(n)={1\over 2 \pi i}\oint \d t P_\infty(t)
t^{-n-1}$. The function $P_\infty(t)$ has a `large' singularity at $t=1$,
but in addition has singularities at all other roots of unity - see \odl\ on
the validity of ignoring these contributions asymptotically. This method was
used by Hardy and Ramanujan to find their celebrated formula, here
given in a less detailed form as,
\eqn\asymppn{
p(n)\sim {1\over 4 n\sqrt{3}}\exp\Big(\pi \sqrt{\ts{2\over 3} n }\Big) \, ,
} which was improved by Rademacher to give $p(n)$ exactly. Their method
relied crucially on the modular properties of $P_\infty(t)$.

Focusing now on the two bosonic fundamental field case for which, in the
large $N$ limit, \recursNnm\ with \gengrelpp\ gives exact counting, at
issue is first finding asymptotic values for the numbers
$Q(n,m,\b)=Q(n,m,b_1,\dots,b_m)$, with constraint equation $\sum_{j=1}^m j
b_j=m$, defined by \eqn\functaexp{ \prod_{j=1}^m{1\over
(1-t^j)^{b_j+1}}\prod_{j>m}{1\over 1-t^j}=1+\sum_{n=1}^\infty Q(n,m,\b) t^n
\,. } Having found these we may then attempt to find the dominant
contribution to \recursNnm\ with \gengrelpp\ for large $N$. In order to
give asymptotic values for $Q(n,m,\b)$
we may follow \gea\ and apply a formula due to Meinardus which gives a 
general result for the generating function \eqn\genmein{ \prod_{n=1}^\infty
(1-t^n)^{-a_n}=1+\sum_{n=1}^\infty r(n) t^n \,. } A detailed version of
Meinardus' theorem may be found in \gea\ but for purposes of brevity we may
note that it implies that, as $n\to \infty$, \eqn\rnform{ r(n)\sim
C\,n^{\kappa} \, \exp\Big(\big(A
\Gamma(\al+1)\zeta(\al+1)n^\al\big)^{1/(\al+1)} (\al+1)/\al \Big) \,, } where
$\zeta(s)=\sum_{j=1}^\infty j^{-s}$ is the Riemann zeta function and the
constants $C, \kappa, \al, A$ are determined by the auxiliary Dirichlet
series, \eqn\auxdir{ D(s)=\sum_{j=1}^\infty {a_j\over j^s} \, , } which must
converge for ${\rm Re}(s)> \al$, a positive real number, and possess an
analytic continuation in the region ${\rm Re}(s)\geq c$, $-1<c<0$, such
that, in this region, $D(s)$ is analytic except at a simple pole at $s=\al$
where it has residue $A$. In terms of $\al,A$ then
\eqn\otherconsts{\eqalign{ C &{} ={1\over \sqrt{2\pi(1{+\al})}}\big(A
\Gamma(\al+1)\zeta(\al+1)\big)^{(1 -2D(0))/2(\al+1)}\exp D'(0) \,,\cr \kappa
&{}=(D(0)-1-{\ts{1\over 2}}\al)/(\al+1) \, . }}

Applying Meinardus' theorem to the case of \functaexp, clearly we have
\eqn\diric{ D(s)=\sum_{j=1}^m {b_j\over j^s}+\zeta(s) \, , } so that,
assuming $m=\sum_{j=1}^m j b_j$ is fixed, $D(s)$ has a simple pole at
$s=\al=1$ where it has residue $A=1$. Using \eqn\vlues{ D(0)=-\half
+\sum_{j=1}^m b_j \, , \qquad \exp D'(0)={1\over \sqrt{ 2
\pi}}\prod_{j=1}^m{1\over j^{b_j}} \, , } then, from \rnform\ with
\otherconsts, we may easily determine that, as $n\to \infty$,
\eqn\asymptotqnb{ Q(n,m,\b)\sim {1\over 4 n \sqrt{3}} \left({\sqrt{6 n}\over
\pi}\right)^{\sum_{j=1}^m b_j}\prod_{j=1}^m {1\over j^{b_j}} \,
\exp\Big(\pi\sqrt{{\ts{2\over 3}} n}\Big) \, . } This reduces to \asymppn\
when $b_j=0$, $1\leq j\leq m$, whereby $Q(n,0,\dots,0)=p(n)$. Using
\recursNnm\ for \gengrelpp\ with \functaexp\ and \asymptotqnb\ then, as
$n\to \infty$, \eqn\expforNnmpo{\eqalign{ \N_{(n,m)} & {}
=\sum_{b_1,\dots,b_m\geq 0\atop \sum_{j=1}^mj b_j=m}Q(n,m,\b)-
\!\!\!\!\sum_{b_1,\dots,b_{m-1}\geq 0\atop \sum_{j=1}^{m{-1}} j b_j=m{-1}}
Q(n{+1},m{-1},\b)\cr &{}\sim {1\over 4 n \sqrt{3}} \left({\sqrt{6 n}\over
\pi}\right)^{m} \, \exp\Big(\pi\sqrt{{\ts{2\over 3}} n}\Big) \,,} } since
$Q(n,m,m,0,\dots,0)$, for $b_1=m$, $b_j=0$, $j>1$, dominates over all other
terms in \expforNnmpo. This gives asymptotic values for
the numbers in \exptwobosc, for counting quarter BPS operators,
transforming in $[m,n{-}m,m]$ $SU(4)_R$ representations,
in the large $N$ limit of free $\N=4$ super Yang Mills,
as previously described.

\noindent \newsec{Counting Operators in the Chiral Ring of $\N=4$
Super Yang Mills}

For the purposes of counting operators in the chiral ring of $\N=4$
super Yang Mills, we denote
corresponding multi-trace partition functions by $\C_{U(N)}(\t)$.

The generating function for $\C_{U(N)}(t)$ for the case of one bosonic
fundamental field has been written in the form \refs{\mald, \bofeng, \bofengt}
\eqn\onebosonicf{ \C(\nu,t)=\prod_{n=0}^\infty{1\over 1-\nu
t^n}=\sum_{N=0}^\infty \nu^N \C_{U(N)}(t) \, , } so that $\nu$ acts as a
chemical potential for the rank of the gauge group $U(N)$. The equivalence
$\C_{U(N)}(t)=\Z_{U(N)}(t)=P_N(t)$, with $\Z_{U(N)}(t)$ as in \compan, is
actually a special case of the $q$-Binomial theorem. Writing - see \andrews\
for notation - \eqn\defqbinpo{ (a;q)_k=(1-a)(1-a q)\cdots (1-a q^{k-1}) \, ,
} then the $q$-Binomial theorem is, for $|x|, |q|<1$, \eqn\qbin{
\sum_{k=0}^\infty {(a;q)_k\over (q;q)_k}\,x^k={(a x;q)_\infty\over
(x;q)_\infty} \, . } (Identifying $\nu=x$ and $q=t$ and setting $a=0$ in
\qbin, so that $1/(\nu;t)_\infty= \C(\nu,t)$ above and $1/(t;t)_N=P_N(t)$ in
\defpnt, then $\C_{U(N)}(t)=\Z_{U(N)}(t)=P_N(t)$ straightforwardly. This
special case of the $q$-Binomial theorem is due to Euler.)

For the two boson case, so that the single particle partition function is
given by $f(t,u)=t+u$ for some $t,u$, then the generating function for the
finite $N$ chiral ring partition function $\C_{U(N)}(t,u)$ is given by
\refs{\mald, \bofeng, \bofengt}
\eqn\onebosonicf{ \C(\nu,t,u)=\prod_{n,m=0}^\infty{1\over 1-\nu t^n
u^m}=\sum_{N=0}^\infty \nu^N \C_{U(N)}(t,u) \, . } This function is more
difficult to analyse in terms of counting though has been investigated by
Stanley \stan\ in relation to partitions - there it has been dubbed the
`double Eulerian' generating function. Through use of the Cauchy-Littlewood
formula, then we may expand $\C_{U(N)}(t,u)$ in terms of partitions of $N$
as, \eqn\expntwov{\C_{U(N)}(t,u)=\sum_{\lambda\vdash
N}h_{\lambda}(t)h_{\lambda}(u) \, ,
}
where 
\eqn\relschuri{
h_\lambda(t)=s_\lambda(1,t,t^2,\dots)\,,
}
so that using an identity for Schur polynomials to be found in \refs{\mac,
\stan}
then
\eqn\expntwovv{
\C_{U(N)}(t,u)=\sum_{\lambda\vdash
N} {\prod_{1\leq i<j\leq N}(1-t^{\lambda_i{-}\lambda_j{+j}{-i}})
(1-u^{\lambda_i{-}\lambda_j{+j}{-i}}) \over \prod_{i=1}^N
(t;t)_{\lambda_i{+N}{-i}}(u;u)_{\lambda_i{+N}{-i}}} \,(t
u)^{\sum_{i=1}^N(i-1)\lambda_i} \, .}
\expntwov\ with \relschuri\ has a natural
interpretation in terms of plane partitions in that, for $\pi$ being all
column-strict plane partitions of shape $\lambda$, $|\pi|=\sum_{i,j}\pi_{ij}$,{\foot{See \stan\ for a
detailed description of plane partitions. Briefly, a column-strict plane
partition of shape $\lambda$ is an array $\pi=(\pi_{ij})$ of non-negative
integers with finitely many non-zero entries, that is arranged in a Young
tableaux with shape $\lambda$ - see appendix A - such that the numbers
$\pi_{ij}$ are weakly decreasing along each row, $\pi_{ij}\geq
\pi_{i\,j{+1}}\geq 0$, and strictly decreasing down each column,
$\pi_{ij}>\pi_{i{+1}\,j}\geq 0$. The sum of the parts of $\pi$ is given by
$|\pi|=\sum_{i,j}\pi_{ij}$. (Note that in contrast to the definition in
\stan, here we are allowing $\pi_{ij}=0$, for some $i,j$, to be a part of
the plane partition $\pi$ with shape $\lambda$.) }} 
\eqn\planep{ h_{\lambda}(t)=s_{\lambda}(1,t,t^2,\dots)=\sum_{\pi}t^{|\pi|}\, . }
Obviously, \expntwov\ with \planep\ generalise for other chiral ring
sectors. (For a different connection between the `double Eulerian'
generating function and major indices of permutations see \stan, p. 385.)
As an illustration of \expntwov\ with \planep, we may consider the case $N=2$
whereby
$\lambda=(2,0),(1,1)$ gives the two possible partitions of $2$.
For $\lambda =(2,0)$ (corresponding to a Young diagram with a
single
row of two boxes) $\pi_{11}\geq
\pi_{12}\geq 0$ gives all column-strict plane partitions of shape $(2,0)$, while for
$\lambda=(1,1)$ (corresponding to a Young
diagram with a single column of two boxes) then $\pi_{11}>\pi_{21}\geq
0$
 gives all column-strict plane
partitions of shape $(1,1)$.
Thus,
\eqn\illustr{\eqalign{
h_{(2,0)}(t)&{}=\sum_{\pi_{11},\pi_{12}\geq 0\atop \pi_{11}\geq
\pi_{12}}t^{\pi_{11}+\pi_{12}}={1\over (1-t)(1-t^2)} \, ,\cr 
h_{(1,1)}(t)&{}=\sum_{\pi_{11},\pi_{21}\geq 0\atop \pi_{11}>
\pi_{21}}t^{\pi_{11}+\pi_{21}}={t\over (1-t)(1-t^2)} \,,}
}
so that, from \expntwov\ for $N=2$,
\eqn\utwocasec{
\C_{U(2)}(t,u)= h_{(2,0)}(t)h_{(2,0)}(u)+h_{(1,1)}(t)h_{(1,1)}(u)={1+u
t\over (1-t)(1-t^2)(1-u)(1-u^2)}\, ,
}
which is the correct result as may be verified by extracting the
$\nu^2$ coefficient in an expansion of \onebosonicf\ up to
$O(\nu^2)$.  

In the large $N$ limit, \eqn\fgtpopo{ \C_{U(\infty)}(t,u)=\prod_{n_1,n_2\geq
0\atop n_1+n_2>0}{1\over 1-t^{n_1} u^{n_2}} \, , } upon which attention is
shortly focused.

For the numbers $\N_{(n,m)}\to {\widehat \N}_{(n,m)}$ counting
quarter BPS primary operators for the chiral ring of $\N=4$ super
Yang Mills, belonging to $[m,n-m,m]$
$SU(4)_R$ $R$-symmetry representations, as in \exptwobosc, 
we have 
{\foot{This formula employs
the orthonormality relation of Schur polynomials described here and has appeared
in a similar context in \char, appendix B.}} \eqn\asuymp{ {\widehat
\N}_{(n,m)}={1\over 8\pi^2}\oint \oint \d t \, \d u \, \C_{U(N)}(t,u)\,
s_{(n,m)}(t^{-1},u^{-1})\,(t^{-1}-u^{-1})^2 \,.}
These may be more conveniently evaluated in terms of
the numbers in \wormhead\ $\M_{(n,m)}\to {\widehat \M}_{(n,m)}$, counting
{\it all} chiral ring quarter BPS operators in the $[m,n{-}m,m]$ $SU(4)_R$
representation, given by
\eqn\suhistra{
{\widehat \M}_{(n,m)}={1\over (2\pi
i)^2}\oint\oint \d t \, \d u \, \, \C_{U(N)}(t,u) \, \,t^{-n-1}u^{-m-1}\,,
}
so that ${\widehat \N}_{(n,m)}={\widehat \M}_{(n,m)}-{\widehat
\M}_{(n+1,m-1)}$.
Defining $\P_{\lambda}(n)$ to be the number of column-strict plane
partitions $\pi$ of shape $\lambda$ so that $|\pi|=\sum_{i,j}\pi_{ij}=n$, then, from
\expntwov\ with \planep\ and \suhistra, 
${\widehat \M}_{(n,m)}=\sum_{\lambda\vdash
N}\P_\lambda(n)\P_\lambda(m)$.  Thus,
\eqn\syhtop{
{\widehat \N}_{(n,m)}=\sum_{\lambda\vdash N}\left
(\P_\lambda(n)\P_\lambda(m)-\P_{\lambda}(n{+}1)
\P_{\lambda}(m{-}1)\right) \, ,
}
counts chiral ring quarter BPS primary operators in $SU(4)_R$
representations $[m,n-m,m]$ for any $n,m$ at
finite
$N$.

\noindent
{\bf Asymptotic Counting for Chiral Ring BPS Operators
at Large $N$}

For asymptotic counting of operators in the chiral ring
of $\N=4$ super Yang Mills at large $N$, a relatively crude
method is employed here which nevertheless captures the exponential
behaviour of counting numbers of interest. This method is based on saddle
point approximations of functions near a dominant singularity - see \odl\
for a useful summary. (Often for physical applications in thermodynamics,
e.g. for entropy formulae, we are interested only in the exponential
behaviour of such numbers anyhow.)

To illustrate, we consider the one boson case in the large $N$ limit
again. We first find a convenient `approximating function' as follows,
\eqn\prodoapp{\eqalign{ P_\infty(t)&{}= \prod_{n=1}^{\infty}{1\over
1-t^n}=\exp\Big(-\sum_{n=1}^{\infty}
\ln(1-t^n)\Big)\cr&
\sim\exp\Big(-\int^\infty_0\d s\,\ln(1-t^s)\Big)=\exp\Big(-{\pi^2\over 6\ln t}\Big)\,,} } which has an
`easier' singularity structure. (The approximation in the second step may be
justified by the Euler-Maclaurin formula for approximating sums by
integrals.) Using \prodoapp\ then for large enough $n$, \eqn\pnform{
p(n)={1\over 2\pi i}\oint \d t \, P_\infty(t)t^{-n-1}\sim {1\over 2 \pi i}
\oint \d t \, e^{g(t)}\,,\qquad g(t)=-{\pi^2\over 6 \ln t}-n \ln t\,. } We
may approximate the latter integral for large $n$ by noting that the
dominant contribution is at the saddle point $t'=e^{-{\pi/\sqrt{6n}}}\sim
1$ for which \eqn\saddle{ g(t')=\pi \sqrt{{\ts {2\over 3}}n}\,,\quad
g'(t')=0\,,\quad g''(t')={2\over \pi }\sqrt{6 n^3}\,e^{\pi\sqrt{2/ 3
n}}=\al\,, } so that, for $t''=t-t'$, \eqn\approx{ p(n)\sim e^{\pi \sqrt{{\ts {2
n/3}}}}{1\over 2 \pi i}\oint \d t''\, e^{{1\over 2}\al t''{}^2}\sim e^{\pi
\sqrt{{\ts {2n/ 3}}}}{1\over 2 \pi}\int^{\infty}_{-\infty}\d s \,
e^{-{1\over 2}\al s^2}= {1\over \sqrt{2\pi \al}}\,e^{\pi \sqrt{{\ts {2n/
3}}}}\,. } Thus, \eqn\asym{ \ln p(n)\sim \pi \sqrt{{\ts{2\over 3}}n}\,, }
which captures the correct behaviour of $\ln p(n)$ for large $n$, according
to \asymppn.

We may proceed analogously for the quarter BPS chiral ring multi-trace
partition function at large $N$, \fgtpopo, which we approximate by
\eqn\prodtwvar{\eqalign{ \C_{U(\infty)}(t,u)& =\prod_{n_1,n_2\geq 0\atop
n_1+n_2>0}{1\over 1-t^{n_1} u^{n_2}}=\exp\Big(-\!\!\! \sum_{n_1,n_2\geq
0\atop n_1+n_2>0}\!\!\!\ln(1-t^{n_1}u^{n_2})\Big)\cr & \sim
\exp\Big(-\int_0^\infty\int_0^\infty\d v\,\d w\,\,\ln(1-t^v u^w)\Big)
=\exp\Big({\zeta(3)\over \ln t\,\ln u}\Big)\,.} }
In this case we have, from \suhistra, \eqn\pnmform{{\widehat \M}_{(n,m)}
\sim {1\over (2 \pi i)^2}\oint \oint \d t \, \d u \, \, e^{g(t,u)}\,,} where
\eqn\expfrg{ g(t,u)={\zeta(3)\over \ln t \, \ln u}-n \ln t-m \ln u\,, } for
$n,m$ large. The dominant contribution to the integral, for $n,m$ both large
{\it and of the same order}, occurs about the point $(t',u')\sim (1,1)$
where \eqn\poit{ t'=e^{-(\zeta(3)m n^{-2})^{1/3}}\, ,\qquad
u'=e^{-(\zeta(3)n m^{-2})^{1/3}} \, , } for which \eqn\saddle{\eqalign{
g(t',u')=3{\root 3 \of{\zeta(3)n m}}\,,\quad {\pr \over \pr
t}g(t,u)\Big|_{(t',u')} ={\pr \over \pr u}g(t,u)\Big|_{(t',u')}=0 \,,\cr
{\pr^2 \over \pr t^2}g(t,u)\Big|_{(t',u')}=2 (\zeta(3)^{-1}n^5
m^{-1})^{1\over 3}\,e^{2(\zeta(3)m n^{-2})^{1/3}}=\al\,,\cr \quad {\pr^2 \over \pr u^2}g(t,u)\Big|_{(t',u')}=2
(\zeta(3)^{-1}m^5 n^{-1})^{1\over 3}\,e^{2(\zeta(3)n m^{-2})^{1/3}}=\be\,,
\cr
{\pr^2 \over \pr t \pr u }g(t,u)\Big|_{(t',u')}= (\zeta(3)^{-1}n^2 
m^2)^{1\over 3}\,e^{(\zeta(3)m n^{-2})^{1/3} +(\zeta(3)n
m^{-2})^{1/3}}=\gamma\,.} } So
long as $m,n$ are both large and of the same order, the saddle point
approximation is justified and we obtain \eqn\pnmasympt{ {\widehat \M}_{(n,m)}\sim
e^{3{\root 3 \of{\zeta(3)n m}}}{1\over 4 \pi^2} \int_{-\infty}^{\infty}
\int^{\infty}_{-\infty}\d v\, \d w\,\,e^{-{1\over 2}(\al v^2+\be w^2+2\gamma v w)}=
h(\al,\be,\gamma)e^{3{\root 3 \of{\zeta(3)n m}}}\,, } where \eqn\determined{
h(\al,\be,\gamma)={1\over 2 \pi\sqrt{\al\be-\gamma^2}}={1\over 2\pi {\sqrt
3}}(\zeta(3)m^{-2}n^{-2})^{1\over 3}\,e^{-(\zeta(3)m
n^{-2})^{1/3}-(\zeta(3)n m^{-2})^{1/ 3}}\,. } We thus have that for $n,m$
both comparably large, \eqn\asump{ \ln {\widehat \M}_{(n,m)}
\sim 3{\root 3 \of{\zeta(3)n
m}}\,, }  so that
\eqn\asymptwobosf{ \ln {\widehat \N}_{(n,m)} =\ln \big(
{\widehat \M}_{(n,m)}-{\widehat \M}_{(n{+}1,m{-}1)}\big)
\sim \ln {\widehat \M}_{(n,m)}\sim 3{\root 3
\of{\zeta(3)n m}}\,. }

It is difficult to check the consistency of this result given the dearth of
literature on these types of multi-variable generating functions and their
asymptotic behaviour, however, we may consider the 
simpler function, also considered in \bofengt,
\eqn\frty{
\C_{U(\infty)}(t,t)=\prod_{n=1}^\infty {1\over (1-t^n)^{n+1}} 
=\sum_{r=0}^\infty \E(r)\,t^r\,,}
where, in terms of the counting numbers ${\widehat \N}_{(n,m)}$, from \exptwobosc,
\eqn\srtfty{\eqalign{
\C_{U(\infty)}(t,t)&{}=
\sum_{n=0}^\infty\sum_{m=0}^n(n{-}m{+1}){\widehat \N}_{(n,m)}
\,t^{n+m} \quad \Rightarrow\quad 
\E(r)=\sum_{m=0}^{[{1\over 2}r]}(r{-}2m{+}1){\widehat \N}_{(r-m,m)}
\, .} }
Hence, 
$\E(r)$ counts quarter BPS primary operators in the chiral ring of
$\N=4$
super Yang Mills, transforming in
$[m,n{-}m,m]$ $SU(4)_R$ representations,
with the same conformal dimensions
$r=n+m$. 
Extracting the
dominant contribution to $\ln \E(r)$ from \srtfty, which occurs at the maximum value
of $m$, $m_{\rm M}=[{1\over 2}r]$, and using \asymptwobosf, 
we obtain
\eqn\ssdftg{
\ln \E(r) \sim \ln {\widehat \N}_{(r-m_{\rm M},m_{\rm M})}
\sim {\ts{3\over 2}}{\root 3
\of{2 \zeta(3)r^2}} \, . }

By considering \frty\ directly, we may employ Meinardus' theorem
(described
in the third section) 
to find the behaviour of $\ln \E(r)$ as $r\to \infty$.
Note, however that
Meinardus' theorem may not be applied directly to \frty\ since the
corresponding auxiliary
Dirichlet series \auxdir, with $a_j=j{+}1$, has two simple poles.
To overcome this difficulty we split \frty\ into a product of
two functions, both separately amenable to application of Meinardus'
theorem.
One is the reciprocal of the Euler function, $P_\infty(t)$ in \prodoapp.
The other, the MacMahon function, is given by
\eqn\macmahon{
M(t)=\prod_{n=1}^\infty{1\over (1-t^n)^n}=\sum_{r=0}^\infty q(r)\, t^r \, ,
}
and has been considered in a similar context as here
in \bofeng.{\foot{The relation of $M(t)$ in 
\macmahon\ to plane partitions is given a description in \gea.
Briefly, $q(r)$ gives the number of ordinary plane partitions $\pi$,
so that $\pi_{ij}\geq \pi_{i{+1}\,j}>0$, $\pi_{ij}\geq
\pi_{i\,j{+1}}>0$,
with $|\pi|=\sum_{i,j}\pi_{ij}=r$.  $p(r)<q(r)$ as ordinary partitions
$\lambda$ are a special case of plane partitions. 
In fact, the
formula for $\ln q(r)$ found here is a special case of a more exact
asymptotic formula first found by Wright \wright\ for the number of plane
partitions $q(r)$ of the number $r$.}}  Writing 
\eqn\anoty{
\C_{U(\infty)}(t,t)=P_\infty(t)M(t) \, ,
}
with $P_\infty(t)$ as in \prodoapp, so  that, using \frty,
\eqn\rtgyu{
\E(r)=\sum_{r_1,r_2\geq 0\atop r_1+r_2=r}p(r_1)q(r_2)
\, ,
}
we may find the asymptotic behaviour of $\ln \E(r)$, as $r\to \infty$,
by extracting the dominant contribution from \rtgyu\
using the asymptotic behaviour of $p(r)$, $q(r)$.  
The auxiliary Dirichlet series for $M(t)$ in \macmahon\
is, from \auxdir\ with $a_j=j$,
\eqn\auxdmac{
D(s)=\zeta(s-1) \, ,
}
which has a simple pole at $s=\al=2$, at which the residue is $A=1$.
Thus, from \rnform,
\eqn\wdief{
\ln q(r)\sim {\ts{3\over 2}}{\root 3
\of{2 \zeta(3)r^2}} \, .}
This is consistent with \ssdftg\ as the dominant contribution to
$\ln \E(r)$ comes from the $r_1=0$ term in \rtgyu\ (since
$p(r)\ll q(r)$ as $r\to \infty$) so that
$\ln \E(r)\sim \ln q(r)$.

It has not escaped attention that the method used here, to
capture the exponential behaviour of asymptotic values for the numbers
${\widehat \M}_{(n,m)}$,
may be easily extended to chiral ring sectors other than the quarter BPS one.  Suppose, for simplicity,
that $Z_j$, $1\leq j\leq k{-1}$,  are commuting bosonic fundamental fields, in the $U(N)$ Lie algebra, so that
the single particle partition function is given by $f(\t)=\sum_{j=1}^{k-1}t_j$, in terms
of the corresponding letters $t_j$.  Let ${\widehat \M}_{(m_1,\dots,m_{k-1})}$
denote the number of independent operators involving products of $m_1$ $Z_1$'s, $m_2$ $Z_2$'s etc. in 
corresponding multi-trace operators.  The multi-trace partition function, in the 
large $N$ limit, is given by,
\eqn\multichbos{
\C_{U(\infty)}(\t)=\prod_{n_1,\dots,n_{k-1}\geq 0\atop n_1+\dots +n_{k-1}>0}
{1\over 1-t_1{}^{n_1}\cdots t_{k-1}{}^{n_{k-1}}}\,,
}
which may be crudely approximated by, similarly as before,{\foot{For ${\rm Li}_n(x)=\sum_{j\geq 1}x^j/j^n$
being the usual Polylogarithm, with
${\rm Li}_{n}(1)=\zeta(n)$, $n> 1$, ${\rm Li}_n(0)=0$, then with the convention ${\rm Li}_1(x)=-\ln(1-x)$,
the following integral 
$$
\int_{0}^1{\d x\over x}\, {\rm Li}_{n}(z x)={\rm Li}_{n+1}(z) \, ,
$$
may be useful for showing this, after a suitable change of variables.}}
\eqn\approxchirbos{
\C_{U(\infty)}(\t)\sim \exp\Big(-\int_0^\infty \prod_{j=1}^{k-1} \d v_j \ln(1-t_1{}^{v_1}\cdots t_{k-1}{}^{v_{k-1}})\Big)= \exp\left({(-1)^{k+1}\zeta(k)\over \ln t_1\cdots \ln t_{k-1}}\right) \,.
}
Thus,
without going into as much detail,
for the analogue of \asump\ we have, (assuming $m_j$ are all comparably large,)
\eqn\flibertygib{
\ln {\widehat \M}_{(m_1,\dots,m_{k-1})}\sim g(t_{1}',\dots,t_{{k-1}}')= k {\root k
\of{\zeta(k)\, m_1 \cdots m_{k-1}}}\, ,
}
where $g(t_{1}',\dots,t_{{k-1}}')$ is the value of
\eqn\valuegyt{
g(t_1,\dots,t_{k-1})={(-1)^{k+1}\zeta(k)\over \ln t_1\cdots \ln t_{k-1}}-m_1 \ln t_1-\dots -m_{k-1} \ln t_{k-1}\, ,
}
at the saddle point $(t_{1}',\dots,t_{{k-1}}')\sim (1, \dots, 1)$, where
\eqn\statpoit{
(\ln t_{1}',\dots,\ln t_{{k-1}}')=-{\root k
\of{\zeta(k)\, m_1 \cdots m_{k-1}}}\,\, ({1/m_1},\dots,{1/m_{k-1}}) \,,
}
so that
\eqn\dervkbos{
{\pr \over \pr t_j}g(t_1,\dots, t_{k-1})\Big|_{(t_{1}',\dots,t_{{k-1}}')}=0\,,\qquad j=1,\dots,k{-1}\,.
}

\flibertygib\ is consistent with a result implied by Meinardus' theorem. The function,{\foot{This may be easily seen from
\multichbos, as the number of solutions to $\sum_{j=1}^{k-1}m_j=n$, where $m_j$ are non-negative
integers, is the binomial number
$\left( {n{+}k{-2}\atop k{-}2}\right)$ which, to leading order in large $n$, behaves like $n^{k-2}/(k{-2})!$.
More properly, we should split the product $\C_{U(\infty)}(t,\dots,t)$ 
into pieces separately amenable to Meinardus' theorem, as for the prior case
for $k=3$,
however, just as for that case, the numbers $c(k,r)$ dominate, and so other contributions are ignored here.
}}
\eqn\dretyg{
\C_{U(\infty)}(t,\dots,t)\sim \prod_{n=1}^\infty (1-t^n)^{-{n^{k-2}/(k-2)!}}=\sum_{r=0}^\infty c(k,r) t^r \,,
}
has auxiliary Dirichlet series, from \auxdir\ with $a_j=j^{k-2}/(k-2)!$,
\eqn\rftyghu{
D(s)={1\over (k-2)!}\,\zeta(s+2-k)\,,
}
which has a simple pole at $s=\al=k{-1}$ at which the residue is $A=1/(k-2)!$, so that, from
\rnform,
\eqn\repoture{
\ln c(k,r)\sim {k\over k{-}1} {\root k \of {(k{-1})\,\zeta(k) \,r^{k-1}}}\,. }
\repoture\ is precisely the result that may be obtained from \flibertygib\
if we maximise the product $m_1\cdots m_{k-1}$, subject to the
constraint $\sum_{j=1}^{k-1} m_j=r$, for which the solution is $m_j=m_{\rm M}=r/(k-1)$
(relaxing the constraint that $m_j$ be non-negative integers, which is irrelevant
asymptotically),
so that  $\ln {\widehat \M}_{(m_{\rm M},\dots,m_{\rm M})}\sim \ln c(k,r)$.

This is applicable to counting multi-trace operators in the eighth BPS chiral ring sector for $\N=4$ super Yang Mills
with fundamental fields $Z,Y,X$ involving $m_1$ $Z$'s, $m_2$ $Y$'s, $m_3$ $X$'s. 
Expanding the corresponding partition function \multichbos, with $k=4$, in terms of Schur polynomials
$s_{(m_1,m_2,m_3)}(\t)$, $m_1\geq m_2\geq m_3\geq 0$, similar to \exptwobosc, 
the expansion coefficients
${\widehat \N}_{(m_1,m_2,m_3)}$ count spinless multi-trace primary
operators transforming in the $[m_2+m_3,m_1-m_2,m_2-m_3]$
$SU(4)_R$ $R$-symmetry representation, with conformal dimensions
$m_1+m_2+m_3$ \char.  Just as in \asymptwobosf, asymptotically
$\ln \N_{(m_1,m_2,m_3)}\sim \ln \M_{(m_1,m_2,m_3)}$.  This counting, however, 
ignores contributions of the fermionic fields
$\lambda, {\bar \lambda}$, which it may be important to include in order
to give correct counting of eighth BPS chiral ring operators.

\noindent
\newsec{Conclusions}

There are some obvious questions not answered by this work. The first is
whether or not the approach in the second section using symmetric
polynomials can give insight into thermodynamics at finite $N$, such as
for the Hagedorn transition, for example. While it gives the large $N$ expression
\infNZ\ in an elementary way, its wider applicability or usefulness to such
questions is unclear. The approach is undoubtedly useful for finding exact
expressions for counting numbers (as in \flambdao, \styo\ and \ndetermined\
for quarter BPS operators) and \flam, \flambosferm\ may be useful for 
analysing counting for more complicated sectors of $\N=4$ super Yang Mills,
with gauge group $U(N)$.

The second question is how the arguments employing symmetric polynomial
techniques here may be extended to other gauge groups, the most pertinent
being perhaps $SU(N)$. Arguments here employing \innero\ and the
orthonormality property of Schur polynomials should remain largely
unaffected for $SU(N)$.  Exact values for counting numbers obtained here
should require some modification for $SU(N)$, though asymptotic values
may be unchanged.

The third question concerns asymptotic values for counting numbers and how 
these may be improved. The asymptotic counting formulae given in such papers
as \refs{\bofeng, \bofengt} for chiral ring sectors are special cases of
formulae such as those of Hardy and Ramanujan, Meinardus, etc., all of which
derive from single variable generating functions. It is hoped that the
expressions \expforNnmpo, \asymptwobosf, \flibertygib, given here for asymptotic counting of
BPS operators, that distinguishes between differing $R$-symmetry
charges, represents a serious attempt at going beyond consideration of
single variable generating functions.{\foot{
After submission of the first version of this paper to the electronic archive, I received an e-mail
from Hai Lin pointing out an interesting comparison between \asymptwobosf\ here
and (2.14) of \Lin, obtained in quite a different context.  The two formulae are
essentially the same given the numerical value $3{\root 3\of{\zeta(3)}}=3.189\dots$,
correct to three decimal places.}}
  Improving upon these formulae will require
more sophisticated techniques, perhaps along the lines used to find
those of Hardy and Ramanujan or Meinardus and employing any
modular properties of the multi-variable functions involved. This issue may
also be important for microscopic counting for Black Holes, as the BPS
solutions found thus far, for $\N=4$ superconformal symmetry, depend on
special values of $R$-symmetry charges \refs{\guto,\gut,\chong,\luc} - see \luct\
for a related detailed discussion.

Thus far, the elegant results for finite $N$ partition functions for chiral
ring sectors have been
interpreted from a largely geometric perspective - it may be interesting to
investigate more how such results are related to the theory of random
matrices and/or symmetric polynomials.

\noindent
{\bf Acknowledgements}

I warmly thank Yang-Hui He, Paul Heslop, Hugh Osborn, Christian
Romelsberger and Christian Saemann for useful comments and
discussions. This work is supported by an IRCSET (Irish Research Council for
Science, Engineering and Technology) Post-doctoral Fellowship.

\vfil \eject \noindent \appendix{A}{Partitions, symmetric group characters,
symmetric polynomials and inner products}

A generic partition $\lambda$ is any finite or infinite sequence
$\lambda=(\lambda_1,\lambda_2,\dots)$ of non-negative integers in decreasing
order $\lambda_1\geq \lambda_2 \geq \dots\geq 0$ containing only finitely
many non-zero terms. Often it is convenient to omit zero entries. The
non-zero entries are called the parts of $\lambda$ the number of which we
denote by $\ell(\lambda)$. The sum of the parts of $\lambda$ is called the
weight of $\lambda$ which we denote by $|\lambda|=\sum_i \lambda_i$. If
$|\lambda|=L$ then $\lambda$ is a partition of $L$ and we write $\lambda
\vdash L$. For convenience we sometimes write $\lambda$ in its frequency
representation which is a reordering of the entries in $\lambda$,
indicating the number of times each successive non-negative integer occurs,
$(1^{a_1},2^{a_2},\dots)$ so that exactly $a_n$ of the parts of $\lambda$
equal $n$ and $|\lambda|=\sum_{n\geq 1} n \, a_n$.

In terms of standard Young diagrams, $\lambda$ corresponds to a Young
diagram of shape $\lambda$, with $\lambda_1$ boxes in the first row,
$\lambda_2$ boxes in the second row etc.; the number of parts
$\ell(\lambda)$ is simply the number of rows and the weight $|\lambda|$ is
the total number of boxes.

For the symmetric group, $\S_N$, the irreducible representations are
labelled by partitions $\lambda\vdash N$ - see \sag\ for a useful summary -
so that, for $X^\lambda(\si)$, $\si\in \S_N$, being a corresponding matrix
representation, then the character of $\si\in \S_N$ in the representation
$X^\lambda$ is $\chi^\lambda(\si)={\rm tr}(X^\lambda(\si))$. The characters
are class functions so that they take a constant value on conjugacy classes
and, recalling that for $\S_N$ the conjugacy classes $K_\mu$ are labelled by
partitions $\mu\vdash N$, corresponding to the cycle structure of a class
representative, then $\chi^\lambda(\si)=\chi^\lambda_\mu$ for all $\si\in
K_{\mu}$. With $z_\lambda$ as defined in \newf, a crucial property of $\S_N$
characters is the orthogonality of the matrix $[z_\mu{}^{-1/2}
\chi^\lambda_\mu]_{\lambda \mu}$. This gives rise to the orthogonality
relations, for $\lambda,\mu\vdash N$, (see also Ch. IV of \lit\ for a
related discussion,) \eqn\orthsym{ {1\over N!}\sum_{\si\in
\S_N}\chi^\lambda(\si)\chi^\mu(\si)=\sum_{\nu\vdash N}{1\over
z_\nu}\chi^\lambda_\nu \, \chi^\mu_\nu =\de_{\lambda\mu}\,, } and
\eqn\rgtyo{ \sum_{\nu\vdash N} \chi^\nu_\lambda \,\chi^\nu_\mu = z_\lambda
\de_{\lambda \mu} \,. }

A convenient basis for $N$ variable symmetric polynomials are Schur
polynomials $s_\lambda(\z)= s_\lambda(z_1,\dots,z_N)$ labelled by
$\lambda=(\lambda_1,\dots,\lambda_N)$. They may be expressed in a number of
ways \refs{\mac,\stan}. For convenience we write them as \eqn\schurply{
s_{\lambda}(\z)=a_{\lambda+\rho}(\z)/a_{\rho}(\z) \,, } where $\rho$, the
Weyl vector, is given by $\rho=(N-1,N-2,\dots,1,0)$ and \eqn\detop{
a_{\lambda+\rho}(\z)=\sum_{\si\in \S_N}{\rm sgn}(\si) \,
z_{\si(1)}{}^{\lambda_1{+}N{-1}}\cdots
z_{\si(j)}{}^{\lambda_j{+N}{-j}}\cdots z_{\si(N)}{}^{\lambda_N} ={\rm
det}[z_i{}^{\lambda_j+N-j}] \, , } with \eqn\vandermond{ a_{\rho}(\z)={\rm
det}[z_i{}^{N-j}]=\prod_{1\leq i< j\leq N}(z_i-z_j)=\De(\x) \, , } being the
Vandermonde determinant. Schur polynomials $s_\lambda(\z)$ have a standard
interpretation as
corresponding to the characters of irreducible $U(N)$ (or, for $\prod_i z_i=1$, 
$SU(N)$) Lie algebra representations. Here, $\lambda$ gives the shape of the
Young tableaux for the corresponding $U(N)$ Lie algebra representation.

For $\lambda=(\lambda_1,\dots,\lambda_N)$ and $\mu=(\mu_1,\dots,\mu_N)$
where $\lambda_i,\mu_i\in \Bbb{Z}$ then, from the definition of \innero\
along with \schurply, \eqn\innerschuro{ \l s_\lambda,s_\mu
\r_N=\sum_{\si\in\S_N}{\rm sgn}(\si)\delta_{\lambda^\si \mu}
=\sum_{\si\in\S_N}{\rm sgn}(\si)\delta_{\lambda \mu^\si}\, , } where, for
any $\lambda'=(\lambda'_1,\dots,\lambda'_N)$,
$\lambda'{}^{\si}=\si(\lambda'+\rho)-\rho$ is the shifted Weyl reflection 
of $\lambda'$ by $\si$, with the action of $\S_N$ on $\lambda'$ being given
by
$\si(\lambda'_1,\dots,\lambda'_N)=(\lambda'_{\si(1)},\dots,\lambda'_{\si(N)}
)$. (Equation \innerschuro\ is a reflection of $s_{\lambda}(\x)={\rm
sgn}(\si)s_{\lambda^\si}(\x)$ for any partition $\lambda$ and $\si\in \S_N$
- note that this property is useful for showing \idsf, \idfg. $\lambda^\si$
has a standard interpretation in terms of $U(N)$ Lie algebra representations
- for the Verma module with dominant integral highest weight having
orthonormal basis labels $\lambda$, $\lambda_1\geq \lambda_2 \geq \dots\geq
\lambda_N\geq 0$, then $\lambda^\si$, for $\si\neq {\rm id}_{\S_N}$, are the
orthonormal basis labels for the highest weights of all invariant
sub-modules. This fact may be exploited to derive the Weyl character formula
\schurply\ for the irreducible $U(N)$ Lie algebra representation with
dominant integral highest weight having orthonormal basis labels $\lambda$,
or, alternatively, Young tableaux of shape $\lambda$.)

When $\lambda, \mu$ are partitions so that $\lambda_1\geq \dots \geq
\lambda_N\geq 0$ and $\mu_1\geq \dots \geq \mu_N\geq 0$ then \innerschuro\
reduces to a well defined inner product, \eqn\innerschur{ \l s_\lambda ,
s_\mu \r_N=\de_{\lambda\mu} \, , } so that in this case the Schur
polynomials are orthonormal. Note that in order that $s_\lambda(\x)$ be
non-zero for some arbitrary partition $\lambda$ then $\ell(\lambda)\leq N$,
so that \innerschur\ is zero for $\ell(\lambda)>N$ or $\ell(\mu)>N$.

Another basis for symmetric polynomials are the power symmetric polynomials,
$p_{\lambda}(\z)$, for $\lambda=(\lambda_1,\dots,\lambda_L)\vdash L$, which
are defined by \eqn\poersymp{
p_\lambda(\z)=p_{\lambda_1}(\z)p_{\lambda_2}(\z)\cdots p_{\lambda_L}(\z) \,
,\qquad p_{n}(\z)=\sum_{i=1}^N z_i{}^n \,. } Note that there is no longer
the restriction that $\ell(\lambda)\leq N$ as for Schur polynomials.

Symmetric group characters may be used to relate the two bases for symmetric
polynomials \refs{\mac,\stan} so that, with the definition of
$z_\lambda$ in \newf,
\eqn\retri{ s_\lambda(\z)=\sum_{\mu\vdash N}{1\over z_\mu}\chi^\lambda_\mu
\, p_\mu(\z) \,, } (a theorem of Frobenius) and, for $\lambda\vdash L$,
\eqn\innersymp{ p_\lambda(\z)=\sum_{\mu\vdash L\atop \ell(\mu)\leq
N}\chi_{\lambda}^\mu \, s_\mu(\z) \, . } (\retri\ with
$\chi_\lambda^{(N)}=1$ for all $\lambda\vdash N$ is useful for obtaining
\kropet.)

Regarding the inner product \innero, then using \innersymp\ along with
\innerschur, we then have that, for $\lambda\vdash L$, $\mu\vdash M$,
\eqn\innerpo{ \l p_\lambda,p_\mu \r_N=\de_{LM}\sum_{\nu\vdash L\atop
\ell(\nu)\leq N} \chi_\lambda^\nu \chi_\mu^\nu \, . } Orthogonality of
symmetric group characters implies, from \rgtyo, that for
$|\lambda|,|\mu|\leq N$ then \innerpo\ simplifies to,
with the definition of $z_\lambda$ in \newf, \eqn\infN{ \l
p_\lambda, p_\mu \r_N=z_\lambda \de_{\lambda \mu}\,. }

\vfil \eject

\appendix{B}{Tables}

\medskip
\vbox{

\hskip 1cm \vbox{\tabskip=0pt \offinterlineskip \halign{& \vrule# &\strut \
\hfil# \cr \noalign{\vskip 3pt} \omit &$\N_{(n)}$\hfil && \ \ \hfil &\omit &
\hfil &\omit & \ \hfil &\omit & \ \hfil &\omit & \ \hfil &\omit & \
\hfil&\omit & \ \hfil\cr \omit & $\quad N^{\dps \, n} $ \hfil &&\quad
${}^{\dps 2}$ \hfil &\omit&\ ${}^{\dps 3}$ \hfil &\omit&\ ${}^{\dps 4}$
\hfil &\omit&\ ${}^{\dps 5}$ \hfil &\omit&\ ${}^{\dps 6}$ \hfil &\omit&\
${}^{\dps 7}$ \hfil &\omit&\ ${}^{\dps 8}$ \hfil &\omit&\ ${}^{\dps 9}$
\hfil &\omit&\ ${}^{\dps 10}$ \hfil &\omit&\ ${}^{\dps 11}$ \hfil \cr
\noalign{\hrule} \omit & \ 1 \ \hfil&& \quad 1 \hfil & \omit & \ $1$ \ \hfil
&\omit &\ $1$ \ \hfil &\omit & \ $1$ \ \hfil &\omit &\ $1$ \ \hfil&\omit & \
$1$ \ \hfil &\omit & \ $1$ \ \hfil &\omit & \ $1$ \ \hfil &\omit & \ $1$ \
\hfil &\omit & \ $1$ \ \hfil\cr \omit & \ 2 \ \hfil&& \quad 2 \hfil & \omit
& \ $2$ \ \hfil &\omit &\ $3$ \ \hfil &\omit & \ $3$ \ \hfil &\omit &\ $4$ \
\hfil&\omit & \ $4$ \ \hfil &\omit & \ $5$ \ \hfil &\omit & \ $5$ \ \hfil
&\omit & \ $6$ \ \hfil &\omit & \ $6$ \ \hfil\cr \omit & \ 3 \ \hfil&& \quad
2 \hfil & \omit & \ $3$ \ \hfil &\omit &\ $4$ \ \hfil &\omit & \ $5$ \ \hfil
&\omit &\ $7$ \ \hfil&\omit & \ $8$ \ \hfil &\omit & \ $10$ \ \hfil&\omit &
\ $12$ \ \hfil
&\omit & \ $14$ \ \hfil&\omit & \ $16$ \ \hfil \cr \omit & \ 4 \ \hfil&& 
\quad 2 \hfil & \omit & \ $3$ \ \hfil &\omit &\ $5$ \ \hfil &\omit & \ $6$ \
\hfil &\omit &\ $9$ \ \hfil&\omit & \ $11$ \ \hfil &\omit & \ $15$ \ \hfil
&\omit & \ $18$ \ \hfil &\omit & \ $23$ \ \hfil&\omit & \ $27$ \ \hfil\cr
\omit & \ 5 \ \hfil&& \quad 2 \hfil & \omit & \ $3$ \ \hfil &\omit &\ $5$ \
\hfil &\omit & \ $7$ \ \hfil &\omit &\ $10$ \ \hfil&\omit & \ $13$ \ \hfil
&\omit & \ $18$ \ \hfil &\omit & \ $23$ \ \hfil &\omit & \ $30$ \
\hfil&\omit & \ $37$ \ \hfil\cr \omit & \ 6 \ \hfil&& \quad 2 \hfil & \omit
& \ $3$ \ \hfil &\omit &\ $5$ \ \hfil &\omit & \ $7$ \ \hfil &\omit &\ $11$
\ \hfil&\omit & \ $14$ \ \hfil &\omit & \ $20$ \ \hfil &\omit & \ $26$ \
\hfil &\omit & \ $35$ \ \hfil&\omit & \ $44$ \ \hfil\cr } }

{\eightpoint
{\parindent 1.5cm{\narrower
\noindent
Numbers of multi-trace half BPS primary operators, with conformal dimension
$n$ and
belonging to $[0,n,0]$ $R$-symmetry representations, for free $\N=4$ SYM with
$U(N)$ gauge group.  (For every $N$ there is one
$[0,0,0]$ and $[0,1,0]$ representation - these are omitted
above.)

}}}}

\vskip 0.5cm

\medskip
\vbox{

\hskip 1cm \vbox{\tabskip=0pt \offinterlineskip \halign{& \vrule# &\strut \
\hfil# \cr \noalign{\vskip 3pt} \omit &$\N_{(n,1)}$\hfil && \ \ \hfil &\omit
& \hfil &\omit & \ \hfil &\omit & \ \hfil &\omit & \ \hfil &\omit & \
\hfil&\omit & \ \hfil\cr \omit & $\quad N^{\dps \, n} $ \hfil &&\quad
${}^{\dps 2}$ \hfil &\omit&\ ${}^{\dps 3}$ \hfil &\omit&\ ${}^{\dps 4}$
\hfil &\omit&\ ${}^{\dps 5}$ \hfil &\omit&\ ${}^{\dps 6}$ \hfil &\omit&\
${}^{\dps 7}$ \hfil &\omit&\ ${}^{\dps 8}$ \hfil &\omit&\ ${}^{\dps 9}$
\hfil &\omit&\ ${}^{\dps 10}$ \hfil &\omit&\ ${}^{\dps 11}$ \hfil \cr
\noalign{\hrule} \omit & \ 2 \ \hfil&& \quad 1 \hfil & \omit & \ $1$ \ \hfil
&\omit &\ $2$ \ \hfil &\omit & \ $2$ \ \hfil &\omit &\ $3$ \ \hfil&\omit & \
$3$ \ \hfil &\omit & \ $4$ \ \hfil &\omit & \ $4$ \ \hfil &\omit & \ $5$ \
\hfil &\omit & \ $5$ \ \hfil\cr \omit & \ 3 \ \hfil&& \quad 1 \hfil & \omit
& \ $2$ \ \hfil &\omit &\ $4$ \ \hfil &\omit & \ $5$ \ \hfil &\omit &\ $8$ \
\hfil&\omit & \ $10$ \ \hfil &\omit & \ $13$ \ \hfil &\omit & \ $16$ \ \hfil
&\omit & \ $20$ \ \hfil &\omit & \ $23$ \ \hfil\cr \omit & \ 4 \ \hfil&&
\quad 1 \hfil & \omit & \ $2$ \ \hfil &\omit &\ $5$ \ \hfil &\omit & \ $7$ \
\hfil &\omit &\ $12$ \ \hfil&\omit & \ $16$ \ \hfil &\omit & \ $23$ \
\hfil&\omit & \ $30$ \ \hfil
&\omit & \ $40$ \ \hfil&\omit & \ $49$ \ \hfil \cr \omit & \ 5 \ \hfil&& 
\quad 1 \hfil & \omit & \ $2$ \ \hfil &\omit &\ $5$ \ \hfil &\omit & \ $8$ \
\hfil &\omit &\ $14$ \ \hfil&\omit & \ $20$ \ \hfil &\omit & \ $30$ \ \hfil
&\omit & \ $41$ \ \hfil &\omit & \ $57$ \ \hfil&\omit & \ $74$ \ \hfil\cr
\omit & \ 6 \ \hfil&& \quad 1 \hfil & \omit & \ $2$ \ \hfil &\omit &\ $5$ \
\hfil &\omit & \ $8$ \ \hfil &\omit &\ $15$ \ \hfil&\omit & \ $22$ \ \hfil
&\omit & \ $34$ \ \hfil &\omit & \ $48$ \ \hfil &\omit & \ $69$ \
\hfil&\omit & \ $92$ \ \hfil\cr \omit & \ 7 \ \hfil&& \quad 1 \hfil & \omit
& \ $2$ \ \hfil &\omit &\ $5$ \ \hfil &\omit & \ $8$ \ \hfil &\omit &\ $15$
\ \hfil&\omit & \ $23$ \ \hfil &\omit & \ $36$ \ \hfil &\omit & \ $52$ \
\hfil &\omit & \ $76$ \ \hfil&\omit & \ $104$ \ \hfil\cr } }

{\eightpoint
{\parindent 1.5cm{\narrower
\noindent
Numbers of multi-trace quarter BPS primary operators, with conformal
dimension $n{+1}$ and
belonging to $[1,n{-1},1]$ $R$-symmetry representations, for free $\N=4$ SYM with
$U(N)$ gauge group. ($n=0,1$ cases are all zero.)

}}}}

\vskip 0.5cm

\medskip
\vbox{

\hskip 1cm \vbox{\tabskip=0pt \offinterlineskip \halign{& \vrule# &\strut \
\hfil# \cr \noalign{\vskip 3pt} \omit &$\N_{(n,2)}$\hfil && \ \ \hfil &\omit
& \hfil &\omit & \ \hfil &\omit & \ \hfil &\omit & \ \hfil &\omit & \
\hfil&\omit & \ \hfil\cr \omit & $\quad N^{\dps \, n} $ \hfil &&\quad
${}^{\dps 2}$ \hfil &\omit&\ ${}^{\dps 3}$ \hfil &\omit&\ ${}^{\dps 4}$
\hfil &\omit&\ ${}^{\dps 5}$ \hfil &\omit&\ ${}^{\dps 6}$ \hfil &\omit&\
${}^{\dps 7}$ \hfil &\omit&\ ${}^{\dps 8}$ \hfil &\omit&\ ${}^{\dps 9}$
\hfil &\omit&\ ${}^{\dps 10}$ \hfil &\omit&\ ${}^{\dps 11}$ \hfil \cr
\noalign{\hrule} \omit & \ 3 \ \hfil&& \quad 3 \hfil & \omit & \ $5$ \ \hfil
&\omit &\ $10$ \ \hfil &\omit & \ $14$ \ \hfil &\omit &\ $21$ \ \hfil&\omit
& \ $27$ \ \hfil &\omit & \ $36$ \ \hfil &\omit & \ $44$ \ \hfil &\omit & \
$55$ \ \hfil &\omit & \ $65$ \ \hfil\cr \omit & \ 4 \ \hfil&& \quad 3 \hfil
& \omit & \ $6$ \ \hfil &\omit &\ $14$ \ \hfil &\omit & \ $21$ \ \hfil
&\omit &\ $36$ \ \hfil&\omit & \ $50$ \ \hfil &\omit & \ $73$ \ \hfil &\omit
& \ $96$ \ \hfil &\omit & \ $130$ \ \hfil &\omit & \ $163$ \ \hfil\cr \omit
& \ 5 \ \hfil&& \quad 3 \hfil & \omit & \ $6$ \ \hfil &\omit &\ $15$ \ \hfil
&\omit & \ $25$ \ \hfil &\omit &\ $44$ \ \hfil&\omit & \ $66$ \ \hfil &\omit
& \ $101$ \ \hfil&\omit & \ $142$ \ \hfil
&\omit & \ $200$ \ \hfil&\omit & \ $267$ \ \hfil \cr \omit & \ 6 \ \hfil&& 
\quad 3 \hfil & \omit & \ $6$ \ \hfil &\omit &\ $15$ \ \hfil &\omit & \ $26$
\ \hfil &\omit &\ $48$ \ \hfil&\omit & \ $74$ \ \hfil &\omit & \ $118$ \
\hfil &\omit & \ $171$ \ \hfil &\omit & \ $251$ \ \hfil&\omit & \ $346$ \
\hfil\cr \omit & \ 7 \ \hfil&& \quad 3 \hfil & \omit & \ $6$ \ \hfil &\omit
&\ $15$ \ \hfil &\omit & \ $26$ \ \hfil &\omit &\ $49$ \ \hfil&\omit & \
$78$ \ \hfil &\omit & \ $126$ \ \hfil &\omit & \ $188$ \ \hfil &\omit & \
$281$ \ \hfil&\omit & \ $398$ \ \hfil\cr \omit & \ 8 \ \hfil&& \quad 3 \hfil
& \omit & \ $6$ \ \hfil &\omit &\ $15$ \ \hfil &\omit & \ $26$ \ \hfil
&\omit &\ $49$ \ \hfil&\omit & \ $79$ \ \hfil &\omit & \ $130$ \ \hfil
&\omit & \ $196$ \ \hfil &\omit & \ $298$ \ \hfil&\omit & \ $428$ \ \hfil\cr
} }

{\eightpoint
{\parindent 1.5cm{\narrower
\noindent
Numbers of multi-trace quarter BPS primary operators, with conformal
dimension $n{+2}$ and
belonging to $[2,n{-2},2]$ $R$-symmetry representations, for 
free $\N=4$ SYM with
$U(N)$ gauge group. ($n=0,1$ cases are all zero.)

}}}}

\listrefs

\bye